\documentclass[12pt,preprint]{aastex}

\begin{document}

%
\def\nid{\noindent}
\def\cen{\centerline}
\def\mvs{\vskip 0.2in}
\def\svs{\vskip 0.1in}
%
\def \ie        {\hbox{\it i.e.,~}}
\def \eg        {\hbox{\it e.g.,~}}
\def \cf        {\hbox{\it cf.~}}
\def \vs        {\hbox{\it vs.~}}
\def \etal      {{\it et al.}}
%
\def \msun      {\ M_{\odot}}
\def \lsun      {\ L_{\odot}}
\def \arcmin    {^\prime}
\def \arcsec    {{^{\prime\prime}}}
\def \mpc       {{\rm\ Mpc}}
\def \kpc       {{\rm\ kpc}}
\def \pc        {{\rm\ pc}}
\def \kms       {\hbox{ km s$^{-1}$}}
\def \hr        {{\rm\ hr}}
\def \yr        {{\rm\ yr}}
\def \myr       {{\rm\ Myr}} 
\def \gyr       {{\rm\ Gyr}} 
\def \dln       {\hbox{\rm d$\,$ln }}
\def \dlog      {\hbox{\rm d$\,$log }}
\def \K         { \hbox{$\,$ K} }
\def \H0        {{\rm\ H_{0}}}
\def \kmsmpc    {{\rm\ km\ s^{-1}\ Mpc^{-1}}}
\def \kev       {{\rm\ keV}}
\def \msol      {{\rm M}_\odot}
\def \h         {\hbox{$\, h$} }
\def \hinv      {\hbox{$\, h^{-1}$} }
\def \ergs      { \hbox{$\,$ erg s$^{-1}$} }
\def \rms       {{\it rms~} }
\def \cgsflux   {{\rm\ erg\ s^{-1}\ cm^{-2}}}
\def \cntflux   {{\rm\ cnts\ s^{-1}\ arcmin^{-2}}} 
\def \cc        {{\rm cm^{-3}}}
\def \mag       {{\rm\ mag}}
\def \magarc    {{\rm\ mag\ arcmin^{-2}}}

\def\arcsecpoint{$''\!,$}
\def\arcminpoint{$'\!.$}
\def\deg{$^{\rm o}$}
\def\ltsim{\raisebox{-.5ex}{$\;\stackrel{<}{\sim}\;$}}
\def\gtsim{\raisebox{-.5ex}{$\;\stackrel{>}{\sim}\;$}}


\slugcomment{Accepted for publication in {\it The Astronomical Journal}}

\shortauthors{Deo, Crenshaw, \& Kraemer}
\shorttitle{Nuclear Morphology of NLS1's}

\title{The Host Galaxies of Narrow-Line Seyfert 1s: Nuclear Dust Morphology and Starburst Rings}

\author{R. P. Deo\altaffilmark{1}, D. M. Crenshaw\altaffilmark{1} \& S. B. Kraemer\altaffilmark{2}}

\altaffiltext{1}{Department of Physics and Astronomy, Georgia State University,
Atlanta, GA 30303; deo@chara.gsu.edu and crenshaw@chara.gsu.edu}
 
\altaffiltext{2}{Catholic University of America, and 
the Exploration of the Universe Division, NASA's Goddard Space Flight 
Center, Code 667, Greenbelt, MD  20771; stiskraemer@yancey.gsfc.nasa.gov.}

\begin{abstract}
We present a study of the nuclear morphology of a sample of narrow- and
broad-line Seyfert 1 galaxies (NLS1's and BLS1's) based on broad-band images
in the \textit{Hubble Space Telescope} archives. In our previous study, we
found that large-scale stellar bars at $> 1\kpc$ from the nucleus are more
common in NLS1's than BLS1's. In this paper we find that NLS1's preferentially
have grand-design dust spirals within $\sim 1\kpc$ of their centers.  We also
find that NLS1's have a higher fraction of nuclear star-forming rings than
BLS1's. We find that many of the morphological differences are due to the
presence or absence of a large-scale stellar bar within the spiral host
galaxy. In general, barred Seyfert 1s tend to have grand-design dust spirals
at their centers, confirming the results of other researchers. The high
fraction of grand-design nuclear dust spirals and stellar nuclear rings
observed in NLS1's host galaxies suggests a means for efficient fueling of
their nuclei to support their high Eddington ratios.
\end{abstract}


\section{Introduction}
Seyfert galaxies \citep{seyfert1943} are the most luminous type of active
galaxy found in the nearby ($z \ltsim 0.1$) universe. They have typical
bolometric luminosities $\sim 10^{43}-10^{45} \ergs$, with $M_{B} > -21.5 +
5\log{h_{0}}$ \citep{sg1983} as the accepted criterion for distinguishing a
Seyfert nucleus from a quasar. Their spectra are dominated by high-ionization
atomic emission lines. \citet{kw1974} showed that there are two distinct
classes of Seyfert galaxy: type 1 Seyferts with broad (FWHM $> 1000 \kms$)
permitted emission lines and superposed narrow (FWHM $\ltsim 500 \kms$)
emission lines from forbidden and permitted transitions; while in type 2
Seyferts broad permitted emission lines are absent.
\citet{oster1977,oster1981} introduced further classifications from 1.2 to 1.9
with numerically larger types having weaker broad-line components relative to
the narrow lines. Using spectropolarimetry, \citet{ma1983} noticed that in NGC
1068 the broad-line region can be seen in polarized light, leading to the idea
that the nature of the central continuum source is similar in both types of
Seyferts.

\citet{op1985} coined the term {\it Narrow-Line Seyfert 1s} (NLS1's) to denote
Seyferts 1s with spectra generally like those of classical Seyfert 1s (strong
Fe~II, [O III] $\lambda 5007$, $\lambda 4959$ relatively weak compared to
hydrogen Balmer series) but with permitted line widths much narrower than
typical Seyfert 1s. \citet{goodrich1989} specified that all NLS1's have
FWHM(H$\beta$) $< 2000\kms$, and this is now the currently accepted criterion
to distinguish NLS1's from BLS1's. The emission characteristics of NLS1's
place them on one extreme of eigenvector 1 of \citet{bg1992}, which was
determined from principal component analysis (PCA) of a large sample of
low-redshift AGN. The PCA analysis confirmed that strong Fe~II, weak [O III]
and narrow H$\beta$ lines are the defining characteristics of the NLS1 class
in the optical regime.

The current widely accepted model for AGN consists of a supermassive black
hole (SMBH) accreting matter via an accretion disk. For NLS1's, the current
paradigm is that NLS1's possess black holes of relatively modest mass ($\leq
10^{7} \msun$) that are being fed at or close to the relative Eddington
accretion rate \citep{pdo1995}. This view is supported by recent observational
results that indicate that NLS1's possess significantly smaller black hole
masses than their broad-line counterparts
\citep{mkc2001,wandel2002,peterson2004}. In this scenario, the narrow-line
widths are simply due to the smaller black hole mass.

In order to fuel AGN, matter must be transported all the way from kiloparsec
scales to the central engine. Thus the matter must lose almost all of its
angular momentum via some process. One possibility involves mergers or tidal
interactions with neighboring galaxies \citep{tt1972, adams1977}. However
studies of environments of Seyfert galaxies show no evidence for a statistical
excess in number of companion galaxies or any recent merger events
\citep{dryh1998} in Seyfert galaxies as compared to normal galaxies. Another
process that has received considerable attention is gas inflow along a
galactic stellar bar \citep{sss1980}. However most observational studies have
found similar fraction of bars in Seyferts and normal galaxies
\citep{heckman1980,sss1980,ho1997,mulchaey1997,mr1997}. Since there is strong
evidence that most normal galaxies contain inactive SMBH \citep{kr1995}, this
indicates that while gas may be transported to the inner kiloparsec via
large-scale stellar bars, there are other factors that contribute to the
presence of nuclear activity.

\citet{shlosman1990} proposed the bar-within-bar fueling scenario, where the
secondary gas/dust bar develops due to non-axisymmetric instabilities in the
gas disk within the inner kiloparsec. \citet{maio2000} found evidence for gas
motion along a secondary bar-like structure within $\sim 100 \pc$ of the
central engine in the Seyfert 2 Circinus galaxy. However it appears that
either gaseous/dusty secondary bars are fairly rare among Seyfert galaxies or
they are relatively small ($\leq 100\pc$) and we are not yet able to see them
due to limited resolution at the distances of most nearby Seyferts (tens of
\mpc). Secondary stellar bars are apparently not that uncommon in nearby
galaxies \citep[see,][]{es2002,es2003}.

Perhaps the most efficient way to detect gaseous inflow to the nucleus is
through the extinction caused by its embedded dust. {\it HST} optical and IR
images indicate that only 10-20\% of Seyferts galaxies show nuclear dust bars
\citep{rm1999, mp1999, pogge2002}. Instead, the {\it HST} images reveal that
most ($\sim 80\%$) Seyfert galaxies show nuclear dust spirals \citep{rm1999,
pogge2002, martini2003a}. However \citet{martini2003b} find that nuclear dust
spirals in Seyfert galaxies are not statistically more numerous that those in
normal galaxies. Thus, it is unclear exactly how the dust spirals fuel the AGN
and what mechanism(s) control the onset of nuclear activity. Although these
structures are common to both barred and unbarred Seyfert galaxies, all of the
``grand-design'' nuclear dust spirals (\ie those with two long symmetric arms)
are found in barred galaxies only \citep{martini2003b}.  These grand-design
nuclear spirals appear to connect to dust lanes on the leading edge of the
large-scale stellar bars.

If the high accretion rate paradigm for NLS1's is correct, it suggests that
the fueling of the AGN is more efficient in NLS1's than in their BLS1
counterparts. In \citet{ckg2003} (hereafter, CKG03), we provided evidence that
NLS1 nuclei are hosted mostly in barred Seyfert galaxies as compared to BLS1
nuclei: 65\% of the NLS1's have bars, while only 25\% of BLS1's have
bars. These large-scale stellar bars typically begin at $\sim 1\kpc$ from the
nucleus and extend to $5-10\kpc$. They represent an efficient means of
transporting large amounts of gas and dust to the inner kiloparsec region,
which can presumably support large accretion rates. However we did not study
the inner regions ($< 1\kpc$) of the BLS1's and NLS1's, where establishing a
connection between Seyfert 1 type and inner morphology would be even more
crucial for understanding the fueling of the active nucleus.

The motivations for this paper came from the need to understand the nuclear
structures that exist within the central regions of NLS1's. There are no
previous studies comparing NLS1 and BLS1 samples for differences in nuclear
morphology. Many of the earlier studies \citep{mp1999,rm1999} were focused
mainly on Seyfert 1.8 to Seyfert 2 galaxies. \citet{pogge2002} and
\citet{martini2003a} included about equal numbers of Seyfert 1's and 2's and
investigate the differences between Seyfert 1's and 2's. We note that in
surveys of nuclear regions of both active and normal galaxies with
\textit{HST} \citep{mp1999,rm1999,pogge2002,martini2003b} and ground-based
imaging studies \citep{es2002}, it was found that nuclear dust spirals are
seen in similar frequencies in \textit{both} active as well as inactive
galaxies. Hence it is not clear if nuclear spirals are indeed efficient in
fueling the central source. Further, \citet{martini2003a} find that barred
galaxies (either active or inactive) preferentially show a grand-design type
of nuclear morphology, while unbarred ones show tightly-wound nuclear
spirals. Since host galaxies of NLS1's are found to be barred from CKG03, we
might expect to see grand-design as the preferred nuclear morphology for
NLS1's.

\section{Sample Selection and Analysis}

\subsection{Sample Selection}
Our sample contains the \textit{HST} broadband (primarily F606W) Wide Field
Planetary Camera 2 (WFPC2) images of Seyfert 1 galaxies obtained in the
snapshot survey of \citet{mgt1998}. This is a uniform sample of 91 Seyfert 1
galaxies with $z \leq 0.039$ derived mainly from the compilation of
\citet{vcv2001}. A typical exposure for each snapshot was 500s, and nearly all
of the galaxies fell on the PC chip, which has a resolution element of
$\sim0.\arcsec 1$. The number of useful NLS1's in this sample is small
(13/91); we excluded MARK 335, as it appears as a point source in the WFPC2
image, leaving 12 NLS1's out of 91 Seyfert 1's. We did not use the five
additional NLS1's at higher redshifts from our previous study (see Table~1 in
CKG03) as we do not have enough resolution at their distances to study nuclear
characteristics. We have also not used UGC 05025 which is a NLS1 with $z$ of
$0.026$ as the F814W exposure time is just 80 secs and the nuclear region is
not well exposed. However we note that this galaxy is barred. Not all galaxies
in the sample had their centers within the PC chip; following is a list of
these:  F1146 (WF2), PKS 0518-458 (WF4), WAS 45 (WF4), UM 614 (WF2). These are
all BLS1's and hence have little effect on our results. We decided to not
include IR~1319-164 in our analysis as most of the galaxy is outside the PC
chip in the WFPC2 field of view on the sawtooth side.

We compiled the redshift (z), axis ratio (b/a), the Hubble stage (T) and the
absolute blue magnitude, $M_{B}^{0}$ from NED for the whole
sample. $M_{B}^{0}$ is computed from the corrected asymptotic blue magnitude
$B_{T}^{0}$ (corrected for interstellar extinction and the K-correction). We
used $B_{T}^{0} - M_{B}^{0} = 5 \log{cz/H_{0}} + 25$, with $H_{0} = 71
\kms\mpc^{-1}$, \citep{spergel2003}. There are galaxies for which we had to
assign a value to the numerical Hubble stage index (T), this value was
assigned by looking at the morphological classification given as part of the
basic data for an object in NED and the table describing the coding of
morphological types in the Third Reference Catalogue of Bright Galaxies
\citep[ hereafter RC3]{1991trcb.book.....D}. We have also used The de
Vaucouleurs Atlas of Galaxies by Buta, Corwin and Odewahn (in preparation) to
assign these values. The values of $B_{T}^{0}$ were selected only from RC3 for
consistency. The axial ratio ($b/a$) were chosen from the RC3 data section
provided in NED.

Table~1 shows our sample and its properties. The first column gives the name
of the galaxy, the second column gives the redshift of the galaxy from CKG03, 
and the third column gives the RC3 axial ratio ($b/a$). The fourth column
gives the numerical Hubble stage index ($T$). The fifth column gives the
corrected absolute blue magnitude ($M_{B}^{0}$) computed as given above.  The
sixth column gives the Seyfert 1 class, \ie whether the galaxy is a NLS1 or a
BLS1. The seventh column gives the large-scale morphology as classified in
this paper. The eighth column gives the nuclear morphology classification for
each galaxy. Three galaxies had main morphology classifications that could not
be defined. These are HEAO 2106-098 (point source), MARK 40 (?), MARK 335
(point source). IR 1319-164 was not included in the analysis as most of the
galaxy is outside the PC field of view. This resulted in a sample of 87
Seyfert 1's where both large-scale and nuclear structures could be classified.

We tested the sample for selection biases that may have been introduced due to
the heterogeneous nature of our sample. Figure~1 shows the histogram plots of
the four host galaxy parameters. The histogram with solid boundary line shows
the BLS1 sample (75/87 objects), while the shaded histogram with dashed
boundry line shows the NLS1 sample (12/87 objects). In Table~2 we list the
representative numbers that describe the sample. The spread ($\sigma$)
reported are standard deviations for the sample in question. 

The sample as a whole (87 Seyfert 1's) have a median Hubble stage index of
1.0, moderately high inclinations of $46.37^o$ and median redshift of
0.024. When we break up the sample into NLS1 and BLS1 classes based on their
$H_{\beta}$ FWHM (done in CKG03), we see that the NLS1's are slightly more
face-on and have a median T of 3.0 as compared to BLS1's which are more
edge-on and have a median T of 1.0. The NLS1's are also 0.73 mag. less
luminous than BLS1's in the median, which is close to the standard deviation
of both groups. All of the differences are smaller than the quoted spreads of
the sample. However, we discuss their possible impact towards the end of \S 3.

\subsection{Analysis}
We retrieved all the images from the \textit{HST} archives, which were
calibrated with the standard HST pipeline. Since this is a snapshot survey, we
only had single frames per galaxy and hence we employed a routine written in
IDL to detect and remove cosmic rays. This routine is written specifically for
handling WFPC2 data and is tuned for the PC chip. The routine scans the input
image for pixels affected by cosmic rays and iteratively discards them,
replacing them with average values of pixels in a box centered on the pixel
being discarded. The routine also estimates the sky background simultaneously
by selecting various peripheral sections of the input image. This option can
be turned off in case the galaxy covers the whole of PC, which is the case for
most images. Various parameters of the routine control the selection of
scanning box size and how to distinguish a faint star from a cosmic ray
hit. We also replaced bright foreground stars within the image with a square
area on the opposite side of the image, with the line joining the two sections
passing through the center of the galaxy. We made sure that none of these
areas contained any dust structures. It should be noted that none of these
foreground stars were close to the centers of the galaxies being studied. This
process was required so that the image enhancement process used would not be
affected by the presence of areas of large intensity apart from the central
point source. Since we are looking for dust structures near the nuclear
source, the cleaner the image, the better the contrast enhancement. Residual
cosmic rays were examined and cleaned by eye using a combination of IRAF and
IDL tools.

We have employed the method used by \citet{pogge2002} to enhance the contrast
of dusty structures. This process has been called ``structure mapping'' and is
based on the Richardson-Lucy (R-L) deconvolution process. The structure map is
the correction image that emerges from the second iteration of an R-L image
reconstruction. It highlights unresolved and marginally resolved structures,
as the first-order smooth structures are removed. One starts with a good
estimate of the \textit{HST} point spread function (for WFPC2 detectors) and
uses it to perform an operation similar to unsharp masking via division. One
divides the original image with a PSF-convolved version of the original image,
multiplying the resulting image with the transpose of the PSF. This results in
an image which contains the high frequency components enhanced in
contrast. The resultant image forms the second-iteration corrector image in a
R-L deconvolution process \citep[see][]{pogge2002}. It is crucial that a
properly matched PSF be used to generate these structure maps.

We used 2-D Gaussian fits to the saturated cores of the Seyfert 1s to
determine the location of the central source on the chip, and then generated
PSFs by using the TinyTim software \citep{kh1999}. The form of the
\textit{HST} WFPC2 PSF depends on the following parameters in the order of
importance: filter used for the observations, the location on WF or PC
detector, the secondary mirror focus position, and the color of object being
observed. The latter two parameters were not of great importance, as the final
structure map did not show significant improvement in quality when these
parameters were tweaked. Most of the galaxy centers were not near any of the
available observed PSFs for WFPC2, hence using TinyTim was the only way to get
reasonable PSFs. PSF subtraction was attempted but yielded inconclusive
results. Information is essentially lost in the saturated core. To attempt
subtraction, the PSFs where scaled to the intensity of the central source and
embedded in image sections the same size as the original image being worked
on. Many images had saturated cores and we estimated best fit 2-D profiles by
looking at 1-D cross-sections through the core and fitting the wings to
estimate the scale factor. The fits were also judged based on the quality of
output structure maps. Since structure maps are quite sensitive to large
variations in brightness levels in pixels, performing PSF subtration and then
forming a structure map from the resulting image was not practical.

Final processed structure maps are shown in Figure~2 (see \citet{mgt1998} for
the original images). The figures are arranged in the same order as the galaxy
name in Table~1. For each galaxy, we show the structure map of a 600 x 600
pixel region of the PC chip which avoids the overscan regions. The first and
third rows show these full structure map sections, while rows two and four
show the nuclear regions of these structure maps zoomed to appropriate size to
facilitate display of nuclear structure. The zoomed nuclear regions for each
galaxy are below the full structure map image. In the figure, each image shows
the size of the region in arcseconds on the vertical axis, as well as the
corresponding projected size of the image in kiloparsec \citep[assuming $H_{0}
= 71 \kms\mpc^{-1}$, see][]{spergel2003} in the top title. The structure map
images are bounded by lower and upper thresholds to display faint structures
properly. The images are also color inverted. Dusty, high extinction areas
appear white, while emission regions appear dark. Compass markers indicate the
North and East directions on the image. The color bar at bottom is provided as
a guide to how the brightness and contrast of the image was stretched between
the applied upper and lower thresholds.

The first two authors independently classified the nuclear structures without
prior knowledge of the Seyfert type (\ie BLS1 or NLS1) or the main galaxy
morphology. We looked at the original image as well as the structure map in
the process. We chose the following notation for our classification: DS for
nuclear dust spiral, DB for nuclear dust bar, A for amorphous dust clouds, DL
for large-scale dust lane passing in front of the central source and ND for no
significant dust structure. We also noticed that several galaxies showed
star-forming rings inside the central kiloparsec and in two cases (MARK 334
and MARK 1044) star-forming nuclear spiral arms. We called the starburst
nuclear spirals, SBS and the nuclear rings, NR. Further, we classified the
dust spirals (DS) into two secondary classes: GD for two arm grand-design
spirals and FL for flocculent multi-armed spirals. The nuclear dust spirals
which could not be classified into these two classes were bunched together
with the notation ``?'' for their secondary classification. We call a nuclear
dust spiral grand-design (GD), if it has two distinct symmetric dust spiral
arms.  Examples in Figure~2 include MARK 1126, MARK 42 and MARK 766. TOL
2327-027 is a spectacular example of this class. The flocculent dust spirals
(FL) were defined to be those that showed more than two distinct spiral arms
peppered with puffy gas and dust clouds. This class essentially bunches
together the classes TW and LW from \citet{martini2003a}. Examples of this
class include:  ESO 323-G77, ESO 354-G4, MARK 1330, MARK 590 and MARK 609. NGC
2639 is a good example of the multi-arm nature of these spirals. There were
other cases with only a single dust arm visible (\eg MCG8-11-11, NGC 6104,
MARK 744), the galaxy has a high inclination which prevented a clean
classification (\eg F51), or showed slightly chaotic grand-design like
structure with a hint of a dust bar-like structure (\eg IC 1816, MARK 334,
UM146). These were not given any special secondary classification and were
grouped together in a category called ``?''. Our class A corresponds to class
C (for chaotic) from \citet{martini2003a}. In the final classification, we
cross-checked and reassigned appropriate classes to the few cases where we
originally disagreed. The Appendix at the end of the text of this paper lists
all the galaxies and the reasons for their individual classifications.

Also during the classification process we noticed that a few of the galaxies
had been previously classified as a spiral (class S) in CKG03, when on
inspection of structure-mapped images and WFPC2 mosaics, they looked to be
barred spirals (class SB). For example, in the PC2 image of ESO 215-G44 in
\citet{mgt1998}, the large-scale bar is not obvious, but can be seen clearly
in a structure map. Other such cases have been recorded in the Appendix. These
galaxies have since been reclassified as SB in Table~1. In CKG03, they had
concluded that excluding point sources, ellipticals, irregular and
unclassified (main morphology class) galaxies (12 out of 97):  34\% (29/85) of
spiral Seyfert 1's are barred, 65\% (11/17) of NLS1's are barred, and 26\%
(18/68) of BLS1's are barred, indicating a high fraction of barred host
galaxies for NLS1's. With our revised classification, we now conclude from the
CKG03 sample of 97 Seyfert 1's that, excluding point sources, ellipticals,
irregular and unclassified (main morphology class) galaxies (12 out of 97), we
now have: 47\% (40/85) of spiral Seyfert 1's are barred, 76\% (13/17) of NLS1
host galaxies are barred and 40\% (27/68) of BLS1 host galaxies are
barred. Thus the incidence of large-scale bars in NLS1's is still much larger
than that in BLS1's.

\section{Results}

In Table~1, we give the results of our classifications.  Column 7 gives the
large-scale morphology based on the structure maps and WFPC2 mosaic images
(see CKG03 for the original classifications). Column 8 gives the nuclear
morphology classification for each galaxy. Within the parenthesis in column 8
is the secondary nuclear classification.  Galaxies that were not given a
formal secondary dust spiral classification are included in the category
``?''.

Table~3 and Table~4 shows the frequencies of these classes as fractions;
number counts for galaxies are given in parenthesis along with one sigma
uncertainities assuming a binomial distribution (since all the classes are
independent and each structure is either present in the galaxy or not). The
distributions are given for the entire sample of 87 Seyfert 1's, as a function
of class (NLS1 \vs BLS1) and as a function of the host galaxy morphology
(barred spirals \vs unbarred spirals). Table~4 shows the distributions for GD,
FL and the undefined (``?'') categories of dust spirals. All of the entries in
this table come from the galaxies that show dust spirals (Table~3, column~3)
as their primary nuclear morphology.

Figure~3 shows the bar plots for each class of nuclear structure. For each
graph, the vertical axis is frequency of the structure and the horizontal axis
has the various classes as in Table~3. On the top of each bar is the fraction
corresponding to the class being presented. The plots on the right side
correspond to comparison of barred vs unbarred galaxies in the sample while
the plots on the left side correspond to NLS1's \vs BLS1's.

We do not see nuclear dust bars, in agreement with \citet{pogge2002}.
\citet{es2002,es2003} demonstrate how stellar secondary bars can be uniquely
identified with help of isophotal analysis and unsharp-masking. Since
structure mapping is similar to unsharp-masking, we could have noticed the
presence of secondary stellar bars, but we did not detect any.

Table~3 and Figure~3 show that $83\%$ of NLS1's and $67\%$ of BLS1's have
nuclear dust spirals, implying that nuclear spirals are the favored
morphological features, in agreement with \citet{martini2003b}. In Table~4, we
see that $80\%$ (8/10) of NLS1's with nuclear dust spirals have grand-design
type nuclear spirals as compared to $32\%$ (16/50) for BLS1's with nuclear
dust spirals. We also see that, $69\%$ (22/32) of barred spirals with nuclear
dust spirals have grand-design structure, compared to $7\%$ (2/28) in the
unbarred spirals. Since the sample of Seyferts with barred galaxy morphology
is more than doubled by adding barred BLS1 (27/75) to the NLS1 (9/12) sample,
and yet the percentage of grand-design nuclear dust spirals (22/32, 69\% of
barred Seyfert 1 sample with dust spirals) remains almost the same as for
NLS1's (8/12, 67\%), we conclude that large-scale stellar bars are the
principal driver of the grand-design dust structure. Even though we have only
12 NLS1's in our sample, 9 are barred and we see that 8 of them show
grand-design nuclear dust spirals. This lends support to the idea that higher
fueling rates in NLS1's are helped by the transfer of gas on kiloparsec scale
via large-scale stellar bars which almost always form grand-design dust
spirals within $1 \kpc$ of the nuclei of NLS1's.

During the classification process for nuclear dust structures, we noticed that
several galaxies in our sample showed nuclear star formation in the form of
stellar nuclear rings or star-forming nuclear spiral arms. ESO 323-G77, IR
1249-131, MARK 42, MARK 493, MARK 530, MARK 744, MARK 896, MARK 1044, NGC
1019, NGC 6212, NGC 7469, TOL 2327-027 and WAS 45 show nuclear star-burst
rings (see Figure~2). MARK 334 and MARK 1044 show nuclear spiral arms with
star forming sites embedded in them. Table~3 and bar plots in Figure~3 show
that $42\%$ ($5/12$) of NLS1's show recent star-formation in nuclear
rings. One out of these (MARK 1044) has star formation in the nuclear spiral
instead of the nuclear ring. In comparison only $12\%$ ($9/75$) BLS1's show
recent nuclear star formation in the form of nuclear rings. Again only one
(MARK 334) shows star formation in the nuclear spiral arms. We do not think
that we have missed any inner stellar rings due to the presence of luminous
point sources, unless they are very small. The size of the saturated point
source is typically much less than $0.\arcsec5$ and at a median $z$ of
$0.024$; $0.\arcsec5$ (about $10$ pixels) corresponds to about $245 \pc$ at
resolution of {\it HST}. We note that all galaxies that host stellar nuclear
rings also show grand-design dust spirals and are barred galaxies; see the NR
category in the top right plot in Figure~3.

As mentioned previously, the NLS1's and BLS1's in our sample show slight
differences in their luminosities ($0.73$ mag.) , inclinations ($11.1^o$) and
Hubble stage ($2$ stages). It is unlikely that these differences have an
impact on our ability to detect the nuclear dust morphology and the presence
of nuclear rings.

As we have discussed, both observational and theoretical studies show that the
presence or abscence of a large-scale stellar bar is the principal driver in
determining the nuclear morphology. Since it is difficult to detect
large-scale bars in highly inclined system, we tested the robustness of our
results by excluding galaxies with inclinations greater than $60^o$. With this
constrain, we have 10 NLS1's and 66 BLS1's. From this reduced sample, 9 out of
10 NLS1's (90\%) have nuclear dust spirals. Out of these 9 dust spiral, 7 are
grand-design (77\%) and one flocculent (11\%). In comparison, out of 66
BLS1's, 48 (73\%) show nuclear dust spirals. 15 (23\%) of these are
grand-designs, while 22 (33\%) are flocculents. Further, 8/10 (80\%) NLS1's
are barred compared to 27/66 (41\%) BLS1's, showing that NLS1's are more
barred than BLS1's as for the whole sample. Of the total 35 barred galaxies in
this reduced sample of 76 galaxies, 21 (60\%) have grand-design nuclear
spirals, while 4 (11\%) are flocculents. Of the 32 unbarred galaxies in this
reduced sample, 19 (59\%) show flocculent nuclear spiral, while one shows
grand-design (3\%). Further, 5/10 (50\%) NLS1's show nuclear rings, while only
9/66 (14\%) of BLS1's show nuclear rings. Thus the difference in inclination
does not seem to contribute to a selection bias for bars.

We noted above that intrinsically NLS1 and BLS1 samples differ in their
absolute blue magnitudes. We suppress the contribution of the nuclear point
source when creating the structure maps. Any residual contribution to the
magnitude due to the Hubble type of the galaxy or the nature of its bulge,
will be very much smaller than the initial luminosity difference present in
the original image between the point source and the rest of the galaxy. Thus
we think our ability to detect faint dusty structures near point sources has
not been affected by the intrinsic magnitude differences between the NLS1 and
BLS1 host galaxies.

Thus in conclusion the trends we see in our original sample of 87 galaxies
seem to be robust against small variations in the host galaxy parameters.

\section{Discussion}
Our statistical analysis shows that the grand-design nuclear dust spirals are
largely present in barred galaxies, which is consistent with previous studies.
We have also found that NLS1 galaxies, which are thought to have relatively
high accretion rates, tend to show large scale stellar bars, nuclear rings and
grand-design nuclear spirals. Are grand-design nuclear spirals indeed more
efficient in fueling the central nucleus?

Here we summarize from the literature the theoretical efforts to answer this
question. The linear density wave theory \citep{gt1978,gt1979} forms the
theoretical underpinnings of hydrodynamical simulations of gas inflow in
barred galaxies. \citet{engshlos2000} showed that the morphology of the
nuclear gas/dust spiral depends on two factors: central mass concentration and
sound speed in the gas (\ie gravitational potential and amount of turbulence
present in the gas disk). The pitch angle (angle between the spiral arm and a
tangent to a circle intersecting with the spiral arm at radius R) of the
spiral is thus dependent on these two factors and ideally one can probe these
factors based on the morphological features of the nuclear disk. We note that
this has not been done so far by any study. In the presence of a large-scale
stellar bar potential, the hydrodynamical simulations from several studies
\citep{engshlos2000, pa2000, jogee2002, macie2002, macie2004a, macie2004b}
show that inflow along leading-edges of the stellar bars form ``grand-design''
type spiral structure in the central kiloparsec. It seems from the discussions
in these papers that the presence of the large-scale stellar bar potential
overwhelms the effects of nuclear gravitational potential and turbulence
throughout most of the central kiloparsec. This effect is reflected in the
tendency of tightly-wound nuclear spirals to avoid barred galaxies in
agreement with our observations (see FL category in bottom right bar plot in
Figure~3). The merging of grand-design nuclear spiral arms with the
leading-edge bar shocks indicates that the nuclear spiral is not decoupled
from the large-scale stellar bar, and that the nuclear spiral pattern is {\it
maintained} by the bar potential.

\citet{macie2002} show that the velocity field of the gas in grand-design
nuclear spirals has a large negative divergence. This is indicative of strong
shocks along the curving dust lanes, which are needed to drive gas down to the
scale of tens of parsec. The merging of grand-design spiral arms with
leading-edges of large-scale bars occurs regardless of underlying nuclear
potential or sound speed in gas, hinting at the importance of large-scale bars
in driving the grand-design nuclear spiral. Simulations of \citet{macie2004b}
indicate that the average inflow rates at $\sim 1\kpc$ are $\sim 0.7 \msun
\yr^{-1}$, at $250 \pc$ they decrease to $\sim 0.2 \msun \yr^{-1}$, and at a
distance of $40 \pc$ they are $\sim 0.03 \msun \yr^{-1}$ in models with both a
central SMBH of $10^{8}\msun$ and a large-scale stellar bar. The inflow at
distances of 40 pc is $\sim 20$ times smaller in models without a black
hole. In the absence of a secondary bar on scales of hundreds of parsec, one
probably needs a strong main bar potential to drive spiral shocks close to
tens of parsec scales. Since local Seyfert galaxies have mass accretion rates
of $\sim 0.01 \msun\yr^{-1}$ \citep{pbook1997}, it seems plausible that the
analysis of \citet{macie2004b} may be in the right direction. Further in the
case of flocculent and tightly-sound nuclear spirals, it is not clear at
present if the inflow rates are substantially reduced as compared to those in
grand-design nuclear spirals, as detailed simulations have not been done.

The presence of strong shocks in the ISM is accompanied by star-formation
activity. We noted in the previous section that we see more nuclear rings in
NLS1's as compared to BLS1's, and the same is true for a barred \vs unbarred
comparison. Presence of these bright stellar nuclear rings indicates that the
central kiloparsec has been fueled in the last few hundred megayears. Further,
bright star-forming regions are often seen on the outer periphery of the main
curving dust lanes in these systems. Studies of the ring phenomena in
galaxies, have shown that rings are resonance phenomena and are often found in
galaxies with a large-scale bar, \citep{buta1996}. The size of nuclear rings
in our sample corresponds with typical values for nuclear rings measured by
other observers \citep{laine2002,es2002,martini2003a}, and ranges from about
$250 \pc$ to $1 \kpc$; TOL 2327-027 would be an exception due to its large
($\approx 5 \kpc$) circumnuclear disk (see Appendix). \citet{laine2002}
mention that size distribution of nuclear rings peaks at about the same radius
as the location of the ILR of the main large-scale bar.

On the whole, we conclude that large-scale stellar bars drive the formation of
and maintain grand-design nuclear dust spirals, which plausibly provide
sufficient inflow rates to fuel the SMBHs in NLS1's. Recent simulations and
the presence of star formation confined in a nuclear ring in the central
kiloparsec give support to the idea that the gas loses its angular momentum
via nuclear spiral shocks driven by the main bar potential. However, it is not
clear what the mass inflow rates in flocculent nuclear spirals are. Since
flocculent spirals are not driven by orbital resonances of a non-axisymmetric
gravitational potential like a bar, it is plausible to expect that spiral
shocks in flocculent spirals are much weaker, leading to less efficient
accretion rates. Observational support for this is that very little star
formation is seen in flocculent type nuclear spirals in unbarred galaxies,
while grand-design spirals show enhanced star formation on the outer edges of
their spiral arms and in stellar nuclear rings \citep[ see also appendix in
this paper]{martini2003a}. However inflow rates in flocculent spiral need to
be studied using numerical simulations.

\section{Conclusions}
We have analyzed {\it HST} broad-band (F606W) images of a sample of 91 Seyfert
1 galaxies to study their nuclear morphology. We employed structure maps to
enhance fine dust structure in the nuclear regions. Accompanying images in
Figure~2 provide a good repository to study nuclear dust structures in Seyfert
1 galaxies.

Our sample contains 12 NLS1 hosts and 75 BLS1 host galaxies. This allowed us
to compare the nuclear morphology of NLS1 host galaxies to the BLS1 host
galaxies. In \citet{ckg2003} (CKG03), we had noticed that NLS1's showed more
large-scale bars as compared to BLS1's. We have revised the main morphological
classification from CKG03 with the aid of structure maps, and conclude that
76\% of NLS1 host galaxies are barred as compared to 40\% for BLS1
hosts. Overall, almost half (47\%) of the spiral Seyfert 1 galaxies from CKG03
are barred.

In this paper, we find that nuclear dust spirals are the most common kind of
morphology in the central kiloparsec of Seyfert 1 galaxies. We find that 80\%
of the NLS1's have grand-design type nuclear spirals as compared to 32\% for
BLS1's. Further we see that 69\% of barred galaxies have grand-design
morphology as compared to 7\% for unbarred ones. This is in agreement with the
trend noted by \citet{martini2003b}. This is also indicative of the fact that
these nuclear spirals are being driven and maintained by large-scale stellar
bars. We also find that most BLS1 host galaxies have multi-arm, flocculent or
chaotic nuclear dust spirals.

We find that 42\% of the NLS1's have nuclear star formation in the form of
nuclear rings as compared to 11\% of BLS1's. Similar distribution is seen when
one compares barred galaxies with the unbarred ones. This again indicates that
the large-scale bar is the main driver of these differences. This strengthens
the idea that large-scale bars are important to support high fueling rates in
NLS1's. Thus our results in this paper support the fueling scenario for barred
\vs unbarred galaxies via nuclear dust spirals. Recent simulations of gas and
dust inflow in barred galaxies give support to the idea that the galactic disk
gas loses its angular momentum via nuclear spiral shocks which are being
driven by the main bar potential within the central kiloparsec.

\acknowledgments

This research has made use of the NASA/IPAC Extragalactic Database (NED) which
is operated by the Jet Propulsion Laboratory, California Institute of
Technology, under contract with the National Aeronautics and Space
Administration. This research has also made use of NASA's Astrophysics Data
System Abstract Service. The data presented in this paper were obtained from
the Multi-mission Archive at the Space Telescope Science Institute (MAST).
Support for MAST for non-HST data is provided by the NASA Office of Space
Science via grant NAG5-7584 and by other grants and contracts. We would also
like to thank Dr. Buta for electronic access to The de Vaucouleurs Atlas of
Galaxies.

\newpage
%
%
\appendix
\section*{APPENDIX\\ 
  Notes on Individual Objects}
\small

Here we provide details on why we chose a particular classification for each
galaxy. We also note the various dust features seen and any other special
comments specific to each galaxy. Some of the galaxies show star-forming
regions/stellar clusters in the circumnuclear region, these appear as small
dark globular regions in the value-inverted images and often are associated
with the dusty regions in the galaxy. Common sites for star formation seem to
be dust lanes along the large-scale bars and in multi-arm loosely-wound
spirals. In case of grand-design spirals most of the star-formation seems to
be restricted to the outer edges of the dust lanes curving in to form the
spiral. Often these form stellar nuclear rings. Many nuclear rings however
seem to be associated with the ``loosely-wound'' type of nuclear spirals. We
feel this is an important differentiator and may be indicative of evolution of
nuclear regions of these galaxies. Galaxies with point source as the primary
morphology classification do not have structure maps included in Figure~2 (see
\citet{mgt1998} for original images).

\begin{description}
\item [ESO 215-G14 (SB:DS:GD, BLS1)] This galaxy has a noticeable bar in the
structure map, but was originally classified as S (unbarred spiral) in
CKG03. The bar is along approximately the east-west direction. Two very faint
dust lanes are seen on leading edges of the bar. The structure map reveals a
GD dust spiral in the center. Northern arm of the spiral is more clearly
visible. The bar structure can be brought out by using a PSF with different
size scale, here we chose the one that showed the inner spiral in clearest
detail.
\item [ESO 323-G77 (SB:DS:FL, BLS1, NR)] Shows a multi-arm flocculent nuclear
spiral along with a nuclear ring of star-burst regions encircling the
spiral. Multiple large-scale dust lanes connect with the nuclear spiral at the
position of the ring. Originally classified as a S (unbarred spiral) in CKG03.
\item [ESO 354-G4 (S:DS:FL, BLS1)] Shows filamentary multi-arm spiral structure
on kiloparsec scale. Toward the center, nuclear spiral structure is smooth
and has distinct puffy gaseous arms with small dust lanes embedded in
them. The nuclear spiral is much smoother than the filamentary nature of the
outer spiral. Spiral arms on north-west side are more distinct.
\item [ESO 362-G18 (S:DS:FL, BLS1)] Shows dust spiral structure in the inner
kpc. Half of the galaxy is in the PC chip while the other half in WF2. Mosaic
image shows star forming regions interspersed with dusty lanes. On the WF2 side
a large-scale dust lane curves in toward the nucleus.
\item [ESO 438-G9 (SB:DS:GD, BLS1)] Large-scale dust lanes are apparent in
main spiral arms as well as along the leading-edges of the bar. Prominent star
forming regions are seen along the bar edges. Toward the center, the dust
lanes curve in to form a grand-design type spiral. The inner regions of the
nuclear spiral itself are lost in the bright saturated core. Both the central
kiloparsec as well as the bar structure shows chaotic dusty regions. Another
galaxy similar to this one is MARK 766.
\item [F 51 (SB:DS:?, BLS1)] This galaxy is highly inclined, with large-scale
dust lanes visible. The central kiloparsec shows chaotic dust structure along
with bright emission to the west and south-west of the nucleus. Star-forming
regions are seen at $\sim$7-8$\arcsec$ from the galaxy center, which may be
the outer co-rotation radius of the bar. It is not clear if the galaxy has a
bar or a warped disk or if this is just a inclination effect.
\item [F 1146 (S:DL:-, BLS1)] This galaxy is too distant from us to resolve
any nuclear features. But it has a large dusty disk ($\approx 8 \kpc$) that is
partially blocking the central AGN.
\item [HEAO 1-0307-730 (SB:ND:-, BLS1)] A prototype barred galaxy with two
distinct spiral arms. It is again too distant to resolve nuclear
structures. However the central 2-3$\kpc$ look devoid of dust.
\item [HEAO 1143-181 (I:A:?, BLS1)] An Irregular galaxy, with emission line
gas filaments visible, however it is too distant to resolve nuclear regions.
\item [HEAO 2106-098 (P:ND:-, BLS1)] Classified in CKG03 as a point source,
it seems this galaxy is probably a SB type, the bar is noticeable in the
structure map. However the galaxy is too distant to see any nuclear structure.
\item [IC 1816 (SB:DS:?, BLS1)] This one shows spectacular dust morphology on
all scales. We see a nuclear dust spiral with two distinct arms that seem to
open as the spiral travels inward eventually forming what looks like a
bar-like structure. Curving dust lanes are prominent. This nuclear spiral has
two distinct arms but is not the prototype GD, hence we have chosen to not
include it in the GD category. The dust structure in the inner 500 pc along the
north-south direction looks very similar to a gaseous/dust bar.
\item [IC 4218 (S:DL:-, BLS1)] Prominent dust lanes are seen in the main
galactic disk. However, the inner 1-2$\kpc$ are devoid of dust and are
smooth. A single dust lane is seen all the way to the nucleus on the west side
of the galactic disk. There is a hint of spiral (main) dust inflow along the
inner side of this dust lane. Overall the galaxy is quite inclined.
\item [IC 4329A (S:DL:-, BLS1)] This is an edge-on galaxy with a large dust
lane crossing the line of sight to the central source. No nuclear structures
are visible.
\item [IR 1249-131 (NGC 4748) (S:DS:GD, NLS1, NR)] The large-scale structure
shows a faint stellar bar extending from NE to SW. A curving dust lane from
the NE side merges with the nuclear ring within the central
1-2$\arcsec$. Inside the nuclear star-burst ring, a two arm nuclear dust
spiral can be seen. The structure of the nuclear region is very similar to
that of IC 1816, but we see a star-burst ring at about the same radius as the
dust spiral arms. The star forming regions and/or emission regions are on the
inside edges of the dust lanes. Much of the rest of the central region seems
to be chaotic. We chose to give it a GD classification as the dust lanes merge
with fainter dust lanes along the large-scale bar.
\item [IR 1319-164 (S:-:-, BLS1)] This galaxy was excluded from analysis as
most of it is outside the PC chip on the sawtooth side of WFPC/2 field of
view.
\item [IR 1333-340 (S:DL:-, BLS1)] This one shows strong dust content within
the central kiloparsec. However it is not clear if there is any spiral
structure, hence we classify this as a DL for dust lane.
\item [MCG 6-26-12 (SB:DS:GD, NLS1)] This is a typical SB galaxy. Even though
the central nuclear region is unresolved, we see curving dust lanes close to
the central saturated core, that connect to the straight dust lanes along the
leading-edges of the large-scale bar. Hence this is classified as a GD.
\item [MCG 8-11-11 (SB:DS:?, BLS1)] This galaxy was classified as a S in CKG03
but is a SB. One can clearly see a dust lane approaching the nucleus on the
north side, eventually curving to the inner nuclear regions. Since we do not
see the second dust lane on the opposite side of the bar, we haven't given it
a GD classification.
\item [MARK 6 (S:DL:-, BLS1)] A strong dust lane passes close to the nucleus,
much of the rest of the galaxy shows little dust structure.
\item [MARK 10 (S:ND:-, BLS1)] Very little dust is noticeable in the inner
kiloparsec. The morphology is similar to IC 4218.
\item [MARK 40 (?:ND:-, BLS1)] This shows very little dust content in the
nucleus, and seems to be an interacting system with a tidal tail. (Not
displayed in Figure~2.)
\item [MARK 42 (SB:DS:GD, NLS1, NR)] A prototype barred galaxy with grand-design
nuclear spiral and a star-burst nuclear ring. Curving dust lanes can be traced
for slightly more than $\pi$ radians.
\item [MARK 50 (S:DS:FL, BLS1)] Outer multi-arm spiral structure is too faint
compared to the central source, but is seen in the structure map. This galaxy
probably hosts a flocculent nuclear spiral. There is not much dust content
visible close to the nucleus. There are hints of winding dust lane features
within $1\arcsec$ of the nucleus.
\item [MARK 79 (SB:DS:GD, BLS1)] Two dust lanes along leading edges of the
large-scale bar, curving toward the nucleus are seen. Nuclear structure is not
prominent, mostly emission line gas filaments are visible.
\item [MARK 279 (S:DS:FL, BLS1)] Multi-arm flocculent nuclear spiral structure
is visible. More dust structures are clearly visible on NW side of the nucleus.
\item [MARK 290 (E:ND:-, BLS1)] No significant dust in the nuclear region is
visible. Mistakenly written as a unbarred spiral in CKG03, this galaxy is
probably elliptical.
\item [MARK 334 (S:DS:?, BLS1, SBS)] Chaotic dust structures are seen on all
scales. The central kiloparsec shows a distinct inverted S shaped spiral. The
region connecting the two arms may also be interpreted as a dust bar. This
galaxy may host a weak large-scale bar approximately along NE-SW line. Much of
the nuclear spiral probably hosts sites of star formation. The spiral arms are
relatively bright compared to other nuclear spirals.
\item [MARK 335 (PG 0003+199) (P:-:-, NLS1)] This galaxy is like a point source,
no nuclear structure can be discerned.
\item [MARK 352 (E:ND:-, BLS1)] We see no significant dust in the nuclear region.
\item [MARK 359 (SB:DS:GD, NLS1)] Shows a very chaotic gaseous and dusty
large-scale bar. The dust lanes are not as straight and clearly demarcated.
\item [MARK 372 (S:DS:FL, BLS1)] Shows a prototypical flocculent multi-arm
nuclear dust spiral very similar to galaxies from class TW of
\citet{martini2003a}
\item [MARK 382 (SB:DS:?, NLS1)] Probably hosts a GD dust spiral, but central
source is too bright to see it clearly. A curving dust lane NW of nucleus,
along leading-edge of the large-scale bar is seen. We choose to not give it a
secondary classification.
\item [MARK 423 (S:DS:FL, BLS1)] This galaxy is probably merging with its
edge-on companion and shows a curious mirror-inverted ``?'' like view. There
is extensive star-formation going on in the disk of the galaxy along with a
disturbed dust morphology. We see chaotic gas structures with dust lanes in
the central kiloparsec. A distinct two arm structure connected by a dust lane
passes through the nucleus.
\item [MARK 471 (SB:DS:GD, BLS1)] We see chaotic dust structures within the
large-scale bar. These dust lanes connect together and eventually curve toward
the center to form a GD type structure. Dust can be traced all the way down to
about $0.2\arcsec$ of the nucleus.
\item [MARK 493 (SB:DS:GD, NLS1, NR)] Another prototypical galaxy with strong
large-scale bar with leading-edge dust lanes feeding a central nuclear ring
and a grand-design spiral toward the center. Also notable in this image is the
presence of multiple dust spiral arms outside the nuclear ring. The nuclear
ring is broken in places where these dust spiral arms connect with inner
structure. These arms probably are the four-armed spiral outside the outer ILR
of the large-scale bar \citep[see,][]{macie2004b}.
\item [MARK 516 (S:DS:FL, BLS1)] We see that the nuclear dust morphology is
chaotic and extensive star-formation is going on. The nuclear region shows two
bright nuclei.
\item [MARK 530 (S:DS:FL, BLS1, NR)] This galaxy shows dust content on all scales,
the nuclear spiral has multiple dust arms and is of flocculent (FL) type. To
the SW of the nucleus just on the outer edge of the curving dust arm, a
star-burst region is prominent.
\item [MARK 543 (S:DS:FL, BLS1)] Shows dust on large scales. Multiple spiral
arms littered with star forming regions are seen. However the central
1-2 $\kpc$ appear to be smooth and devoid of dust.
\item [MARK 590 (S:DS:FL, BLS1)] This galaxy is similar to MARK 543, but this
time the central regions are much more clearly visible. One can see multiple
dust spiral arms intermingled with puffy-looking smooth emission. It is at
about the same distance as MARK 543, but the star forming regions and multiple
dust arms are located on the outer edges of the galactic disk as compared to
MARK 543.
\item [MARK 595 (S:DS:FL, BLS1)] Shows a multi-arm flocculent dust spiral. One
also sees a nuclear emission ring-like structure close to the central source
crossed by a straight dust lane, which connects on one side with a dusty
spiral arm. Spiral pattern north of the nucleus is not seen.
\item [MARK 609 (S:DS:FL, BLS1)] Shows multiple dust spiral arms on all
scales. The overall look is that of a chaotic spiral.
\item [MARK 699 (E:ND:-, NLS1)] This galaxy is like a point source and almost
no structure is visible.
\item [MARK 704 (SB:A:?, BLS1)] This is a SB galaxy with very little dust
structure visible due to the bright nucleus, however there is dust structure
that appears to be curving dust lanes just near the point source.
\item [MARK 744 (S:DS:?, BLS1, NR)] This galaxy shows a dusty and gaseous nuclear
spiral, large dust lanes are seen on Northern side of the galactic
disk. Central spiral has multiple dusty arms that wind by more than $2\pi$. A
nuclear ring may be forming at about $1\arcsec$ distance from the nucleus.
\item [MARK 766 (SB:DS:GD, NLS1)] This is a prototype barred NLS1 with large
amount of dust in the large-scale bar along with the curving dust lanes
toward the center, that eventually form a grand-design type of nuclear
spiral.
\item [MARK 817 (SB:DS:GD, BLS1)] Another prototypical barred galaxy with
leading-edge dust lanes culminating in grand-design nuclear spiral.
\item [MARK 833 (I:A:?, BLS1)] This is an irregular galaxy with a hint of
formation of spiral structure and a bar along its NW direction. The dust
structure is mostly amorphous.
\item [MARK 871 (S:DL:-, BLS1)] This is a spiral galaxy with large dust lanes
and a smooth inner disk. A large dust lane is traveling all the way to the
central source on the NW side of nucleus. Similar dust lane is absent on the
SE side, however dust lanes seem to be curving toward the nucleus, about
$6.5\arcsec$ from the nucleus.
\item [MARK 885 (SB:DS:GD, BLS1)] This is a barred galaxy, with leading-edge
dust lanes along the bar. However they do not form a typical GD structure in
the central kiloparsec. The dust lanes south of the nucleus seems to merge
partly with a dust lane from the other side of the bar on the NE side of the
nucleus. The curved dust structure on the north side of the nucleus extends
till it merges toward the nucleus.
\item [MARK 896 (SB:DS:GD, NLS1, NR)] This is another prototype barred galaxy
showing grand-design nuclear spiral with a nuclear ring. Originally classified
as a S (unbarred spiral) in CKG03.
\item [MARK 915 (S:DS:GD, BLS1)] This is a spectacular example of a dust lane
tracing all the way to the nucleus. The galaxy also shows cone-shaped NLR
emission regions which are almost perpendicular to the inflowing dust
structures. The galaxy is highly inclined.
\item [MARK 1040 (S:DL:-, NLS1)] Another spectacular example of a large-scale
dust lane obscuring the central regions. Not much can be inferred from the
nuclear regions.
\item [MARK 1044 (SB:DS:FL, NLS1, SBS)] This is a interesting galaxy showing
large-scale dust lanes along leading-edges of the main bar, feeding what looks
like a multi-arm flocculent type spiral. However the dust lanes can be traced
for about $2\pi$ radians before they disappear in the central saturated PSF
core. The multi-arm stellar spiral could be the four-armed spiral that forms
at the outer ILR. Originally classified as a S (unbarred spiral) in CKG03.
\item [MARK 1126 (SB:DS:GD, BLS1)] This is another prototype barred galaxy
with a distinct grand-design nuclear dust spiral in the center. The inner
extension of the spiral may be interpreted as a dust bar. However the change
in pitch angle of the spiral is quite strong and it seems similar change on
the west side of nucleus is being masked by the strong emission from the
nucleus in that direction. Originally classified as a S (unbarred spiral) in
CKG03.
\item [MARK 1218 (SB:DS:GD, BLS1)] This is a flocculent spiral that is being
fed by large-scale dust lanes. In the nuclear regions, the dust structure is
mostly chaotic. The galaxy is probably barred.
\item [MARK 1330 (SB:DS:GD, BLS1)] Shows a gaseous and dusty nuclear
spiral. The single spiral arm is spectacular and traveling all the way to the
nucleus, where the flow seems to disintegrate in to chaotic gas and dust
clouds. We earlier classified it as a FL based only on the PC image, but after
looking at the mosaic image one can see that the dust extensions on the north
and south side of the nuclear spiral connect to the dust lanes on leading
edges of the large-scale bar. Hence we reclassified it as a GD. Originally
classified as a S (unbarred spiral) in CKG03.
\item [MARK 1376 (S:DL:-, BLS1)] This galaxy is highly inclined and hence it
is impossible to see the nuclear dust structure. An ionization cone like
structure is emanating from the nucleus almost perpendicular to the dust
lanes. Originally not classified (category ?) in CKG03.
\item [MARK 1400 (S:DS:?, BLS1)] The central structure of MARK 1400 is smooth
and very little dust structure is seen. There is hint of a dust spiral in the
central 2-3$\kpc$.
\item [MARK 1469 (S:DL:-, BLS1)] This galaxy is highly inclined with only
large-scale dust lanes visible.
\item [MS 1110+2210 (E:ND:-, BLS1)] This galaxy is completely featureless, with
no detectable dust structure close to the nucleus.
\item [NGC 235 (S:DS:FL, BLS1)] This galaxy shows a flocculent tightly winding
nuclear dust spiral.
\item [NGC 526A (I:DL:-, BLS1)] This is peculiar galaxy with a large veil-like
dust structure in front of the central nucleus. We classified this as a dust
lane type (DL).
\item [NGC 1019 (SB:DS:GD, BLS1, NR)] This is a spectacular galaxy showing two
large-scale spiral arms with dust lanes, a large-scale bar with leading-edge
dust lanes, a distinct nuclear star-forming ring at about $1 \kpc$ from the
nucleus. There is a dust lane traveling all the way to the nucleus from the SE
side inside the nuclear ring. The inner disk appears featureless and mostly
devoid of dust except for the faint dust lanes visible on North and SE of the
nucleus. The SE lane connects with curving dust lane of the northern arm of
the large-scale bar. Faint multi-arm spiral extensions are seen around the
nuclear ring, these are probably similar to the four-armed type spiral
patterns seen in other barred galaxies with an outer ILR.
\item [NGC 1566 (S:DS:FL, BLS1)] This is a Sbc galaxy having a flocculent
dusty nuclear spiral. Very close to the nucleus the spiral shows a curious
inverted S shaped structure in the inner $1\arcsec$.
\item [NGC 2639 (S:DS:FL, BLS1)] This is a spectacular galaxy showing multi-arm
flocculent dust spiral structure on large-scales. In the inner $5\arcsec$ from
the nucleus the disk becomes much smoother with two dust lanes traveling
toward the nucleus. The puffy nuclear gas disk probably has 3 or 4 spiral
arms.
\item [NGC 3227 (SB:DL:-, BLS1)] The inner disk is highly inclined and hence
we can only see dust lanes crossing it on the SW side. Looking at the WFPC2
mosaic image, these dust lanes are the inner parts of the leading-edge dust
lanes of the large-scale bar. The dust within the large-scale bar is also
chaotic. We are not seeing the other edge of the bar on the SE side in the
WFPC2 image. Emission line gas is seen near the nucleus and perpendicular to
the disk plane. Since we cannot see the nuclear structure clearly, we choose
to classify this as a DL. Originally not classified (category ?) in CKG03.
\item [NGC 3516 (SB:DL:-, BLS1)] This is a spiral galaxy with a large chaotic
dust lane traveling toward the nucleus. There are more dusty regions south of
the nucleus. The large-scale structure is not clearly seen, but it seems like
the galaxy is a SB type in the WFPC1 mosaic image, and the bar may lie on
NW-SE line with respect to the nucleus. It is unclear what is generating the
large-scale chaotic dust lane. Originally classified as a S (unbarred spiral)
in CKG03.
\item [NGC 3783 (SB:DS:FL, BLS1)] This is a typical barred galaxy, with a
large-scale star+dust ring at outer co-rotational radius of the bar. The dust
lanes along the bar edges are faint. We can see curving dust lanes from the
bar edges, but apart from that there are multiple dust lanes, so we choose to
go with the FL category for this galaxy.
\item [NGC 4051 (SB:DS:GD, NLS1)] This shows a gas and dust rich nuclear
region. We choose to classify this galaxy as a GD based on the dust lane NE of
the nucleus which connects to the large-scale leading-edge dust lane of the
bar. This is not immediately apparent from the PC image due to proximity of the
galaxy.
\item [NGC 4235 (S:DL:-, BLS1)] This galaxy is highly inclined and we only see
the large-scale dust lane.
\item [NGC 5252 (S:DS:?, BLS1)] This is a curious galaxy with dust filaments
stretching in arcs all the way out to $15\arcsec$ and a stellar bar like
structure which is completely devoid of dust, stretching along the NS
line. One sees a tightly wounding spiral structure NW of the nucleus. Since we
could not properly classify this we chose to only give it the primary
classification of DS.
\item [NGC 5548 (S:DS:FL, BLS1)] This one shows a inwardly winding nuclear dust
spiral. The overall morphology is similar to NGC 7213.
\item [NGC 5674 (SB:DS:GD, BLS1)] This is a prototype barred spiral with
spectacular grand-design nuclear spiral. Dust content is seen on all
scales.
\item [NGC 5940 (SB:DS:GD, BLS1)] This is another barred spiral with a
spectacular grand-design nuclear spiral. Dust content is again prominent on
all scales.
\item [NGC 6104 (SB:DS:?, BLS1)] Dust content is prominent on all scales. A
strong large-scale bar is visible, but only one leading-edge dust lane is
visible on the NE side of the nucleus. This lane curves and moves toward the
nucleus, however since we do not see the other lane we choose to not give it a
secondary classification. This is probably a GD.
\item [NGC 6212 (S:DS:FL, BLS1, NR)] This is a spectacular example of a
multi-arm flocculent spiral structure for the outer disk and a tightly wound
multi-arm structure for the nuclear disk. This is what would be a prototype
class TW nuclear spiral from the classification of \citet{martini2003a}. It
also shows a nuclear star-forming ring at between $2$ and $3 \arcsec$ from the
nucleus.
\item [NGC 6860 (S:DS:FL, BLS1)] This galaxy is partly out of the PC chip, but
one can see the dust lanes that connect the outer dusty disk with the
nucleus. Most of the nuclear disk is smooth.
\item [NGC 7213 (S:DS:FL, BLS1)] This a prototype galaxy for a multi-arm
flocculent inwardly winding nuclear dust spiral. The dust spiral can be traced
all the way to the nucleus. The nuclear disk is packed with large quantities
of gas and dust.
\item [NGC 7314 (S:DS:FL, BLS1)] The inclined nuclear disk shows chaotic dust
content and hence is classified as FL.
\item [NGC 7469 (S:DS:?, BLS1, NR)] Shows a good example of a multi-arm
nuclear spiral with a nuclear star-burst ring.
\item [II SZ 10 (S:ND:-, BLS1)] This is a barred spiral but is too distant to
search for nuclear structures. This was mistakenly written as II ZW 10 in
CKG03 as the MAST fits header gives 'IIZW10' as the value of 'TARGNAME'
parameter.
\item [PKS 0518-458 (E:ND:-, BLS1)] This is a elliptical galaxy but too distant to look for nuclear structures.
\item [TOL 1059+105 (S:ND:-, BLS1)] This galaxy has bright emission line
filaments near its nucleus, but shows no dust content.
\item [TOL 2327-027 (SB:DS:GD, BLS1, NR)] This is a spectacular example of a
nuclear grand-design spiral. It can be questioned if we can call the spiral
that starts at $\approx 5 \kpc$ a nuclear spiral. We can trace winding dust
lanes all the way down to 200 pc from the nucleus. At about $\approx 2 \kpc$
the winding of large dust lanes halts and the galaxy seems to have formed a
gaseous disk; small dust lanes continue through this disk toward the
nucleus. All the star forming regions seem to form in a ring-like structure
and are on the outer side of the curving dust lanes. Originally classified as
a S (unbarred spiral) in CKG03.
\item [UM 146 (S:DS:?, BLS1)] The dust structure in this galaxy is quite faint
in the nucleus which is dominated by the optical emission of the nuclear gas
disk. One can see small dust lanes crossing the disk. It is possible that the
disk is being fueled via large-scale dust lanes, but these are not clearly
visible, except on the east side of the nucleus.
\item [UGC 3223 (SB:DS:FL, BLS1)] This one shows a flocculent multi-arm disk
in the nuclear region. Dust is distributed on large scale in the galaxy and
one can see a dust lane traveling toward the nucleus from the north side of
the galaxy. This dust lane eventually curves to become the nuclear
spiral. There is a ring-like structure around the nucleus. Originally
classified as a S (unbarred spiral) in CKG03.
\item [WAS45 (SB:DS:GD, BLS1, NR)] This galaxy probably has a grand-design
nuclear spiral and is similar in morphology to MARK 1044. A nuclear stellar
ring is clearly visible.
\item [UGC 10683B (SB:DS:?, BLS1)] This one shows a dusty tightly wound nuclear
spiral.
\item [UGC 12138 (SB:DS:GD, BLS1)] This one shows a grand-design nuclear dust
spiral. The bar is probably not strong as the leading-edge dust lanes are not
straight lines but curved.
\item [UM 614 (S:ND:-, BLS1)] Very little dust is seen in the nuclear regions
of this galaxy.
\item [X 0459+034 (E:A:-, BLS1)] This galaxy shows extended emission line gas
filaments but no dust structure is detectable in nuclear region.
\end{description}
\normalsize

\newpage

\clearpage

%
%
\newpage
\figcaption[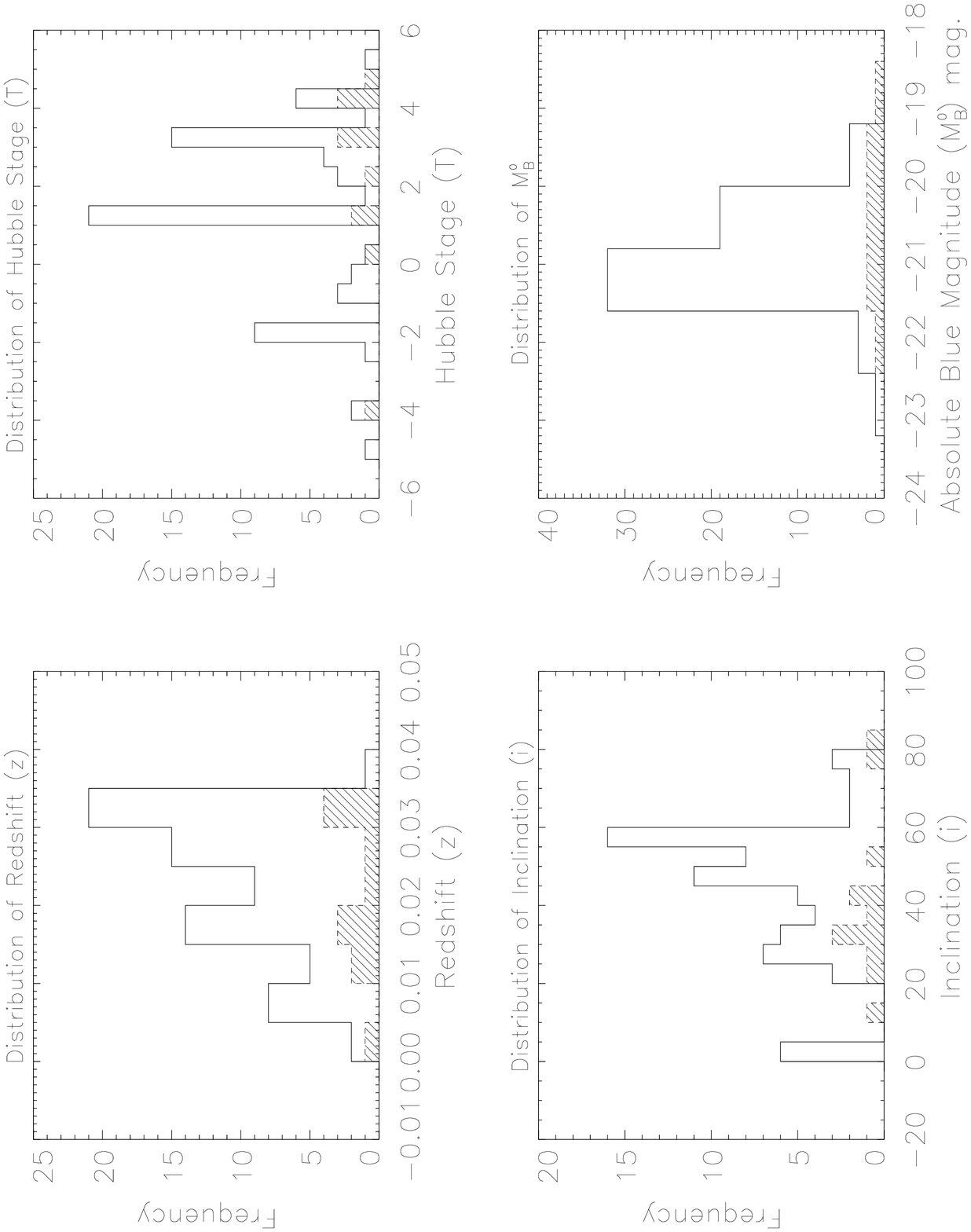]{Host Galaxy Properties of the Sample:  Histograms of
four host galaxy parameters: the redshift (z), the galaxy inclination (i) in
degrees, numerical Hubble stage index (T) and the absolute blue magnitude
($M_{B}^{0}$) for the NLS1 (shaded histogram) \vs BLS1 sample.}

\figcaption[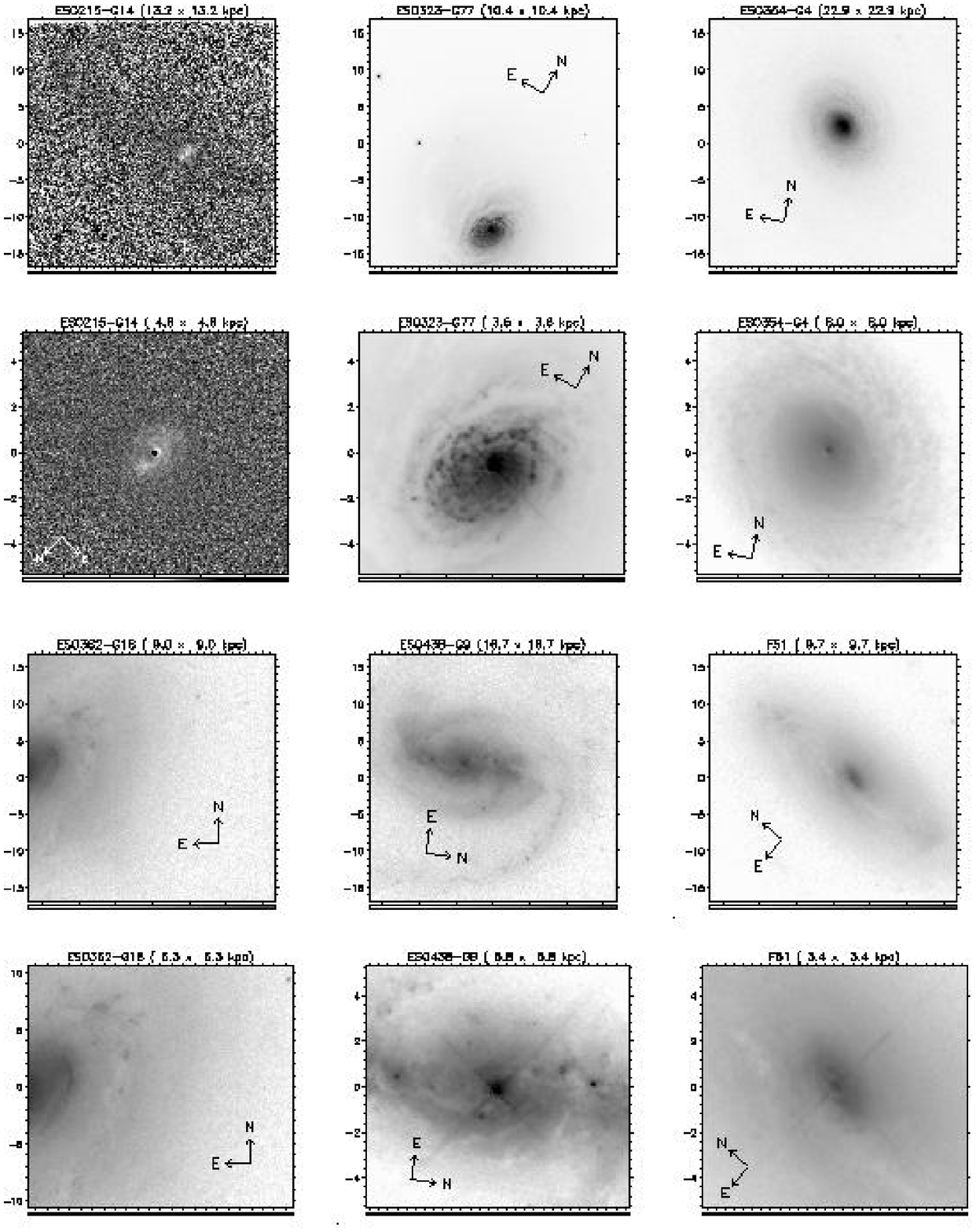]{Structure maps of \textit{HST} WFPC2 images of Seyferts
1s in Table~1:  All images are archival F606W snapshots. Images shown in this
paper are structure maps (see \citet{mgt1998} for original images). Rows one
and three show a 600 x 600 pixel region of the PC chip to avoid overscan
regions. Rows two and four show the same image, but zoomed to show the
nuclear structure. See the scale on the left side of plots for exact
dimensions in arcsecond. All images are color inverted. Dust structures are white
and bright emission regions are dark.}

\figcaption[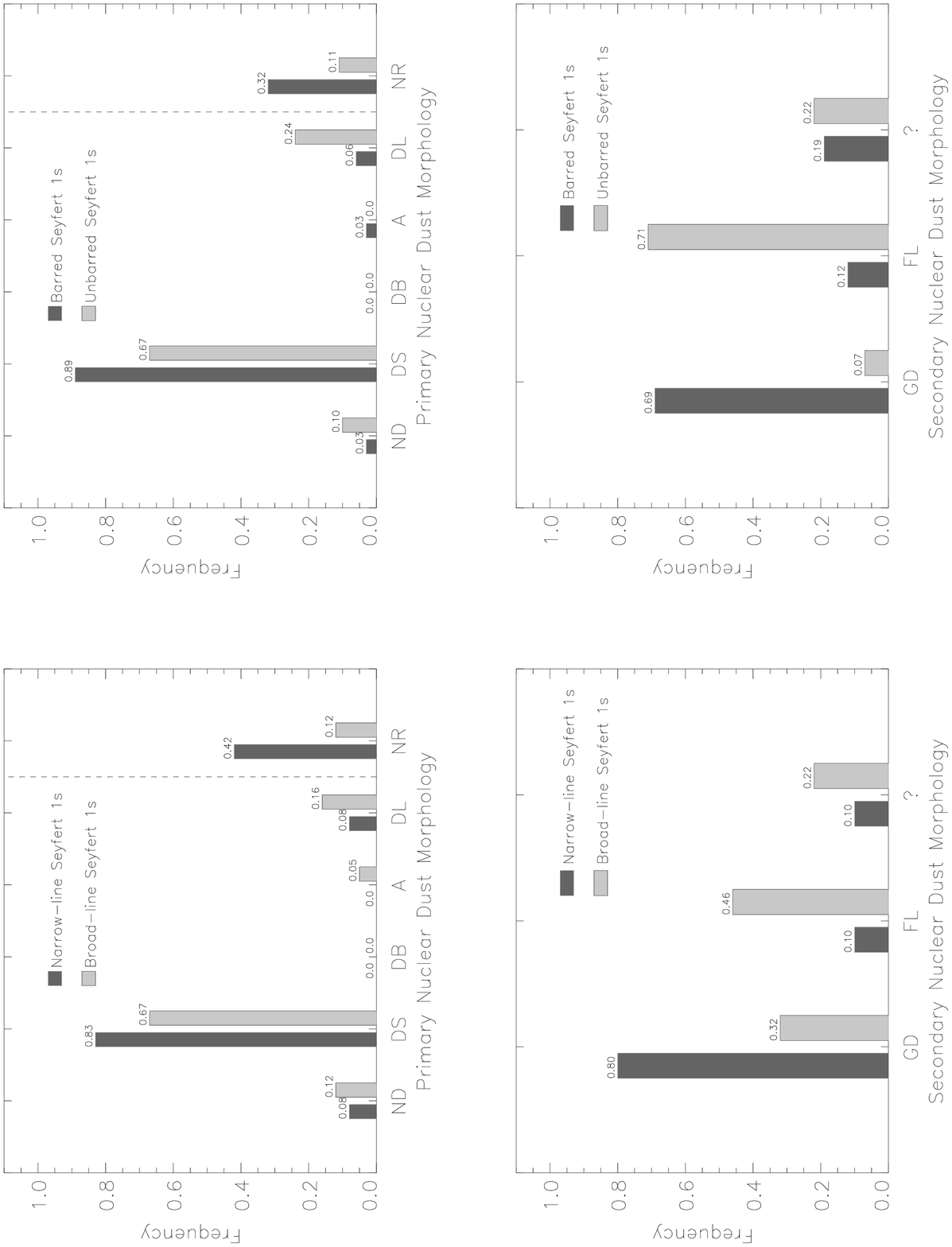]{Frequency of Nuclear Dust Structures: Figure shows bar
plots for various morphological classes shown in Table~3. See \S 2.2 in
text for description of morphology classes. Plots on the right side compare
barred galaxies against unbarred ones. Plots on the left side compare NLS1's
with BLS1's. Plots on the top row pertain to the primary nuclear morphology
classification, while plots on the bottom row pertain to the secondary
classification within the DS class from the top row.  In the top row plots,
the dashed partition separates stellar nuclear rings from nuclear dust
structures. The statistics for nuclear rings also includes two starburst
nuclear spirals: MARK 334 and MARK 1044, for simplicity.}
\clearpage

\newpage
%
%
\begin{deluxetable}{lrrrrcll}
\tablecolumns{8}
\scriptsize
\tablecaption{Nuclear Morphology and Host Galaxy Parameters for Seyfert 1 Galaxies}
\tablewidth{0pt}
\tablehead{
  \colhead{Name} &
  \colhead{Redshift} &
  \colhead{$b/a$$^h$} &
  \colhead{Hubble$^h$} &
  \colhead{$M_{B}^{0}$$^h$} &
  \colhead{Seyfert 1} &
  \colhead{Large-scale} &
  \colhead{Nuclear} \\
  \colhead{} &
  \colhead{} &
  \colhead{} &
  \colhead{Stage} &
  \colhead{} &
  \colhead{Class} &
  \colhead{Morph.} &
  \colhead{Morph.} \\
  \colhead{(1)} &
  \colhead{(2)} &
  \colhead{(3)} &
  \colhead{(4)} &
  \colhead{(5)} &
  \colhead{(6)} &
  \colhead{(7)} &
  \colhead{(8)} 
}
\startdata
ESO 215-G14       & 0.019 & 0.70 &  3.0 &        & BLS1 & SB & DS (GD)      \\
ESO 323-G77       & 0.015 & 0.67 & -2.2 & -20.92 & BLS1 & SB & DS (FL), NR  \\
ESO 354-G4        & 0.033 & 0.82 &  2.7 & -21.53 & BLS1 & S  & DS (FL)      \\
ESO 362-G18       & 0.013 & 0.67 & -0.3 & -20.12 & BLS1 & S  & DS (FL)      \\
ESO 438-G9        & 0.024 & 0.70 &  2.2 & -21.41 & BLS1 & SB & DS (GD)      \\
F 51              & 0.014 & 0.56 &  2.8 & -20.20 & BLS1 & SB & DS (?)       \\
F 1146$^b$        & 0.032 & 0.62 &  3.0 &        & BLS1 & S  & DL           \\
HEAO 1-0307-730   & 0.028 & 0.50 &  1.0 &        & BLS1 & SB & ND           \\
HEAO 1143-181     & 0.033 & 0.88 & 90.0 &        & BLS1 & I  & A            \\
HEAO 2106-099$^d$ & 0.027 & 0.60 &  0.0 &        & BLS1 & P  & ND           \\
IC 1816           & 0.017 & 0.86 &  2.4 & -20.62 & BLS1 & SB & DS (?)       \\
IC 4218           & 0.019 & 0.23 &  1.0 & -20.89 & BLS1 & S  & DL           \\
IC 4329A          & 0.016 & 0.29 & -0.7 & -20.51 & BLS1 & S  & DL           \\
IR 1249-131$^f$   & 0.014 & 0.58 &  3.0 &        & NLS1 & S  & DS (GD), NR  \\
IR 1319-164$^e$   & 0.017 & 0.80 &  3.0 &        & BLS1 & S  & -            \\
IR 1333-340       & 0.008 & 0.60 & -2.0 & -19.24 & BLS1 & S  & DL           \\
MCG 6-26-12       & 0.032 & 0.13 &  4.7 &        & NLS1 & SB & DS (GD)      \\
MCG 8-11-11       & 0.020 & 0.71 &  1.0 & -21.54 & BLS1 & SB & DS (?)       \\
MARK 6            & 0.019 & 0.63 & -0.5 & -20.05 & BLS1 & S  & DL           \\
MARK 10           & 0.030 & 0.39 &  3.0 & -22.62 & BLS1 & S  & ND           \\
MARK 40           & 0.020 & 0.43 & -2.0 &        & BLS1 & ?  & ND           \\
MARK 42           & 0.024 & 0.98 &  2.0 & -20.00 & NLS1 & SB & DS (GD), NR  \\
MARK 50           & 0.023 & 0.60 & -2.0 & -20.05 & BLS1 & S  & DS (FL)      \\
MARK 79           & 0.022 & 1.00 &  3.0 & -21.52 & BLS1 & SB & DS (GD)      \\
MARK 279          & 0.031 & 0.56 & -2.0 & -21.15 & BLS1 & S  & DS (FL)      \\
MARK 290          & 0.029 & 0.89 & -5.0 & -20.37 & BLS1 & E  & ND           \\
MARK 334          & 0.022 & 0.70 & 99.0 & -20.83 & BLS1 & S  & DS (?), SBS  \\
MARK 335$^d$      & 0.025 & 1.00 &  0.0 &        & NLS1 & P  & -            \\
MARK 352          & 0.015 & 0.50 & -2.0 & -19.52 & BLS1 & E  & ND           \\
MARK 359          & 0.017 & 0.83 &  0.0 & -20.44 & NLS1 & SB & DS (GD)      \\
MARK 372          & 0.031 & 0.80 &  1.0 & -21.23 & BLS1 & S  & DS (FL)      \\
MARK 382          & 0.034 & 0.92 &  4.0 & -20.94 & NLS1 & SB & DS (?)       \\
MARK 423          & 0.032 & 0.56 &  3.0 & -21.01 & BLS1 & S  & DS (FL)      \\
MARK 471          & 0.034 & 0.67 &  1.0 & -21.57 & BLS1 & SB & DS (GD)      \\
\tablebreak
MARK 493          & 0.031 & 0.83 &  3.0 & -21.13 & NLS1 & SB & DS (GD), NR  \\
MARK 516          & 0.028 & 0.83 &  4.0 & -20.36 & BLS1 & S  & DS (FL)      \\
MARK 530          & 0.029 & 0.67 &  3.0 & -21.84 & BLS1 & S  & DS (FL), NR  \\
MARK 543          & 0.026 & 1.00 &  6.0 & -20.23 & BLS1 & S  & DS (FL)      \\
MARK 590          & 0.027 & 0.91 &  1.3 & -21.61 & BLS1 & S  & DS (FL)      \\
MARK 595          & 0.028 & 0.68 &  1.0 & -21.00 & BLS1 & S  & DS (FL)      \\
MARK 609          & 0.032 & 0.90 &  1.0 &        & BLS1 & S  & DS (FL)      \\
MARK 699$^d$      & 0.034 & 0.87 & -4.0 & -20.16 & NLS1 & E  & ND           \\
MARK 704          & 0.029 & 0.57 &  1.0 & -20.55 & BLS1 & SB & A            \\
MARK 744          & 0.010 & 0.59 &  1.0 & -20.16 & BLS1 & S  & DS (?), NR   \\
MARK 766          & 0.012 & 0.80 &  1.0 & -19.73 & NLS1 & SB & DS (GD)      \\
MARK 817          & 0.033 & 1.00 &  4.0 & -21.36 & BLS1 & SB & DS (GD)      \\
MARK 833          & 0.039 & 0.75 & 90.0 &        & BLS1 & I  & A            \\
MARK 871          & 0.034 & 0.50 &  0.0 & -21.29 & BLS1 & S  & DL           \\
MARK 885          & 0.026 & 0.57 &  3.0 & -20.41 & BLS1 & SB & DS (GD)      \\
MARK 896          & 0.027 & 0.73 &  3.0 &        & NLS1 & SB & DS (GD), NR  \\
MARK 915          & 0.025 & 0.30 &  3.0 &        & BLS1 & S  & DS (GD)      \\
MARK 1040         & 0.016 & 0.21 &  4.0 & -21.92 & NLS1 & S  & DL           \\
MARK 1044         & 0.016 & 0.86 &  1.0 &        & NLS1 & SB & DS (FL), SBS \\
MARK 1126         & 0.010 & 1.00 &  1.0 &        & BLS1 & SB & DS (GD)      \\
MARK 1218         & 0.028 & 0.50 &  3.0 & -20.83 & BLS1 & SB & DS (GD)      \\
MARK 1330         & 0.009 & 0.74 &  3.0 & -21.47 & BLS1 & SB & DS (GD)      \\
MARK 1376         & 0.007 & 0.24 &  1.0 & -20.09 & BLS1 & S  & DL           \\
MARK 1400         & 0.029 & 0.50 &  1.0 &        & BLS1 & S  & DS (?)       \\
MARK 1469         & 0.031 & 0.36 &  3.0 &        & BLS1 & S  & DL           \\
MS 1110+2210      & 0.030 & 0.86 & -4.0 &        & BLS1 & E  & ND           \\
NGC 235           & 0.022 & 0.54 & -2.0 & -20.87 & BLS1 & S  & DS (FL)      \\
NGC 526A          & 0.018 & 0.53 & -2.0 &        & BLS1 & I  & DL           \\
NGC 1019          & 0.024 & 0.90 &  3.5 & -20.86 & BLS1 & SB & DS (GD), NR  \\
NGC 1566          & 0.004 & 0.80 &  4.0 & -20.98 & BLS1 & S  & DS (FL)      \\
NGC 2639          & 0.011 & 0.61 &  1.0 & -21.14 & BLS1 & S  & DS (FL)      \\
NGC 3227          & 0.003 & 0.67 &  1.0 & -19.33 & BLS1 & SB & DL           \\
NGC 3516          & 0.009 & 0.77 & -2.0 & -20.76 & BLS1 & SB & DL           \\
NGC 3783          & 0.009 & 0.89 &  1.5 & -20.86 & BLS1 & SB & DS (FL)      \\
NGC 4051          & 0.002 & 0.75 &  4.0 & -18.89 & NLS1 & SB & DS (GD)      \\
\tablebreak
NGC 4235          & 0.007 & 0.21 &  1.0 & -20.46 & BLS1 & S  & DL           \\
NGC 5252          & 0.022 & 0.56 & -2.0 & -20.91 & BLS1 & S  & DS (?)       \\
NGC 5548          & 0.017 & 0.93 &  0.0 & -21.47 & BLS1 & S  & DS (FL)      \\
NGC 5674          & 0.025 & 0.91 &  5.0 & -21.55 & BLS1 & SB & DS (GD)      \\
NGC 5940          & 0.033 & 1.00 &  2.0 & -21.60 & BLS1 & SB & DS (GD)      \\
NGC 6104          & 0.028 & 0.82 &  3.0 & -21.39 & BLS1 & SB & DS (?)       \\
NGC 6212          & 0.030 & 0.76 &  3.0 & -20.58 & BLS1 & S  & DS (FL), NR  \\
NGC 6860          & 0.015 & 0.62 &  2.6 & -20.80 & BLS1 & S  & DS (FL)      \\
NGC 7213          & 0.006 & 0.90 &  1.0 & -20.89 & BLS1 & S  & DS (FL)      \\
NGC 7314          & 0.006 & 0.46 &  4.0 & -20.93 & BLS1 & S  & DS (FL)      \\
NGC 7469          & 0.017 & 0.73 &  1.0 & -21.64 & BLS1 & S  & DS (?), NR   \\
II SZ 10$^g$      & 0.034 & 0.60 &  4.0 &        & BLS1 & S  & ND           \\
PKS 0518-458$^c$  & 0.034 & 0.69 & -2.0 & -19.84 & BLS1 & E  & ND           \\
TOL 1059+105      & 0.034 & 0.45 & -1.0 & -20.66 & BLS1 & S  & ND           \\
TOL 2327-027      & 0.033 & 0.68 &  2.7 &        & BLS1 & SB & DS (GD), NR  \\
UM 146            & 0.017 & 0.77 &  3.0 & -20.42 & BLS1 & S  & DS (?)       \\
UGC 3223          & 0.018 & 0.57 &  1.0 & -21.32 & BLS1 & SB & DS (FL)      \\
WAS 45$^c$        & 0.024 & 1.00 &  3.0 & -20.92 & BLS1 & SB & DS (GD), NR  \\
UGC 10683B        & 0.031 & 0.50 &  1.0 &        & BLS1 & SB & DS (?)       \\
UGC 12138         & 0.025 & 0.88 &  1.0 & -21.33 & BLS1 & SB & DS (GD)      \\
UM 614$^b$        & 0.033 & 0.52 & -1.0 & -20.88 & BLS1 & S  & ND           \\
X 0459+034$^a$    & 0.016 & 0.82 & -4.0 &        & BLS1 & E  & A            
\enddata
\tablecomments{All images except as noted were taken with F606W filter.
Large-scale galactic morphology classes are; S: spiral, SB: barred spiral, 
E: elliptical, I: irregular, P: point source and ?: not classified. 
Primary nuclear morphology classes are; ND: no significant dust, DS: dust spiral, 
DB: dust bar, A: amorphous structures, DL: large-scale dust lane, 
NR: Nuclear Ring and SBS: Star-burst (nuclear) Spiral.
Secondary nuclear morphology classes are; GD: grand-design spiral, 
FL: flocculent multi-arm spiral and ?: for no secondary classification.}
\tablenotetext{a}{Observed with the F814W filter.}\tablenotetext{b}{Galaxy positioned in WF2 chip}
\tablenotetext{c}{Galaxy positioned in WF4 chip}\tablenotetext{d}{Not included in Figure~2.}
\tablenotetext{e}{Excluded from analysis, see \S 2.1.}\tablenotetext{f}{Other name: NGC4748}
\tablenotetext{g}{The MAST fits header entry for this galaxy mentions it as II ZW 10 which is incorrect.}
\tablenotetext{h}{Data values were selected from NED for these columns. Some values for the
Hubble stage T were reassigned using classifications from The de Vaucouleurs
Atlas of Galaxies, Buta, Corwin and Odewahn (in preparation).}

\normalsize
\end{deluxetable}
\clearpage

%
%
\newpage
\begin{deluxetable}{lrrrrrrrrr}
\centering
\tablecolumns{16}
\tablecaption{Host Galaxy Properties of the Sample}
\tabletypesize{\footnotesize}
\tablewidth{0pt}
\tablehead{
  \colhead{} &
  \multicolumn{3}{c}{All (87)} &
  \multicolumn{3}{c}{NLS1 (12/87)} &
  \multicolumn{3}{c}{BLS1 (75/87)} \\ 
  \colhead{Parameter} &
  \colhead{Median} &
  \colhead{Mean} &
  \colhead{$\sigma$} &
  \colhead{Median} &
  \colhead{Mean} &
  \colhead{$\sigma$} &
  \colhead{Median} &
  \colhead{Mean} &
  \colhead{$\sigma$}
}
\startdata
Redshift ($z$)          & 0.024  & 0.022  & 0.01  &   0.024 & 0.022  & 0.01  & 0.024  & 0.023  & 0.01  \\
Inclination $i$ (deg.)  & 46.37  & 43.70  & 19.36 &  36.87  & 41.58  & 20.92 & 47.93  & 44.31  & 18.87 \\
Hubble Stage (T)        &   1.0  &   1.3  &  2.3  &   3.0   & 2.1    & 2.4   & 1.0    & 1.2    & 2.3   \\
$M_{B}^{0}$ (mag.)      & -20.88 & -20.80 &  0.69 & -20.16  & -20.40 & 0.93  & -20.89 & -20.86 & 0.64  \\
\enddata
\end{deluxetable}
\clearpage

\newpage
\begin{deluxetable}{lllllll}
\rotate
\centering
\tablecolumns{7}
\tablecaption{Frequency of Primary Nuclear Dust Structures$^a$}
\tabletypesize{\footnotesize}
\tablewidth{0pt}
\tablehead{
  \colhead{Galaxy} &
  \colhead{No} &
  \colhead{Dust} &
  \colhead{Dust} &
  \colhead{Amorph.} &
  \colhead{Dust} &
  \colhead{Nuclear$^d$} \\
  \colhead{Class} &
  \colhead{Dust} &
  \colhead{Spiral$^c$} &
  \colhead{Bar} &
  \colhead{} &
  \colhead{Lane} &
  \colhead{Ring}
}
\startdata
Seyfert 1s (87)$^b$ & $0.11$ ($10 \pm 3.0$)  & $0.69$ ($60 \pm 4.3$) & $0.00$ ($0$) & $0.05$ ($4 \pm 2.0$)  & $0.15$ ($13 \pm 3.3$) & $0.16$ ($14 \pm 3.4$) \\
NLS1's (12)         & $0.08$ ($ 1 \pm 1.0$)  & $0.83$ ($10 \pm 1.3$) & $0.00$ ($0$) & $0.00$ ($0$)          & $0.08$ ($ 1 \pm 1.0$) & $0.42$ ($ 5 \pm 1.7$) \\
BLS1's (75)         & $0.12$ ($ 9 \pm 2.8$)  & $0.67$ ($50 \pm 4.1$) & $0.00$ ($0$) & $0.05$ ($4 \pm 2.0$)  & $0.16$ ($12 \pm 3.2$) & $0.12$ ($ 9 \pm 2.8$) \\
Barred Spirals (36) & $0.03$ ($ 1 \pm 1.0$)  & $0.89$ ($32 \pm 1.9$) & $0.00$ ($0$) & $0.03$ ($1 \pm 1.0$)  & $0.06$ ($ 2 \pm 1.4$) & $0.32$ ($ 8 \pm 2.5$) \\
Spirals (42)        & $0.10$ ($ 4 \pm 1.9$)  & $0.67$ ($28 \pm 3.1$) & $0.00$ ($0$) & $0.00$ ($0$)          & $0.24$ ($10 \pm 2.8$) & $0.11$ ($ 6 \pm 2.3$) \\
\enddata
\tablenotetext{a}{Given as fractions; number of galaxies in each group is given in parenthesis, along with binomial errors.}
\tablenotetext{b}{Out of a total of 91 Seyferts, 87 have both nuclear and large-scale morphological type defined.}
\tablenotetext{c}{Dust spirals are further classified into Grand Design and Flocculent types, see Table~4.}
\tablenotetext{d}{The nuclear ring category also includes the two nuclear spirals showing starburst spiral arms, MARK 334 and MARK 1044, for simplicity.}
\end{deluxetable}
\normalsize
\clearpage

%
%
\newpage
\begin{deluxetable}{llll}
\centering
\tablecolumns{4}
\tablecaption{Frequency of Nuclear Dust Spiral Morphology$^a$}
\tablewidth{0pt}
\tablehead{
  \colhead{} &
  \multicolumn{3}{c}{Nuclear Spiral Morphology} \\
  \colhead{Galaxy Class$^c$} &
  \colhead{Grand Design$^b$} &
  \colhead{Flocculent$^b$} &
  \colhead{Undefined$^b$}
}
\startdata
Seyfert 1s (60)     &  $0.40$ ($24 \pm 4.0$) & $0.40$ ($24 \pm 3.8$) & $0.20$ ($12 \pm 3.1$)  \\
NLS1's (10)         &  $0.80$ ($ 8 \pm 1.3$) & $0.10$ ($ 1 \pm 1.0$) & $0.10$ ($ 1 \pm 1.0$)  \\
BLS1's (50)         &  $0.32$ ($16 \pm 3.3$) & $0.46$ ($23 \pm 3.5$) & $0.22$ ($11 \pm 2.9$)  \\
Barred Spirals (32) &  $0.69$ ($22 \pm 2.6$) & $0.12$ ($ 4 \pm 1.9$) & $0.19$ ($ 6 \pm 2.2$)  \\
Spirals (28)        &  $0.07$ ($ 2 \pm 1.4$) & $0.71$ ($20 \pm 2.4$) & $0.21$ ($ 6 \pm 2.2$)  \\
\enddata
\tablenotetext{a}{Given as fractions; number of galaxies in each group is given in parenthesis, along with binomial errors.}
\tablenotetext{b}{Nuclear spiral classification: these are galaxies that show dust spirals as their primary nuclear morphology.}
\tablenotetext{c}{The numbers in paranthesis in this column come from column 3 in Table~3, \eg out of 87 Seyfert 1's in Table~3, 60 have dust spirals.}
\end{deluxetable}
\normalsize
\clearpage

\newpage
%
%
\begin{figure}[p]
  \setcounter{figure}{0}
  \centering
  \plotone{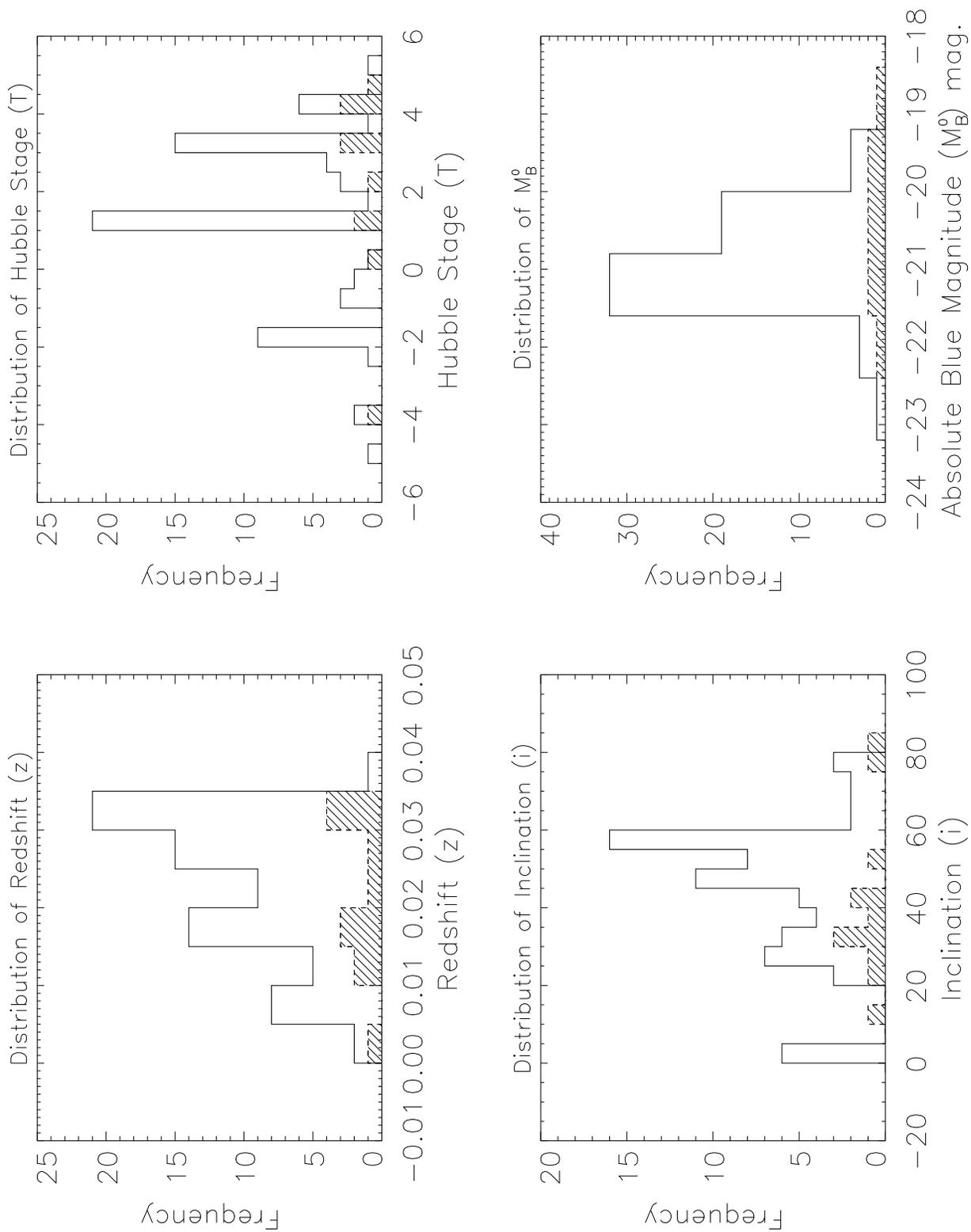}
  \caption{Host Galaxy Properties of the Sample}
\end{figure}
\clearpage

\newpage
%
%
\begin{figure}[p]
  \setcounter{figure}{1}
  \centering
  \plotone{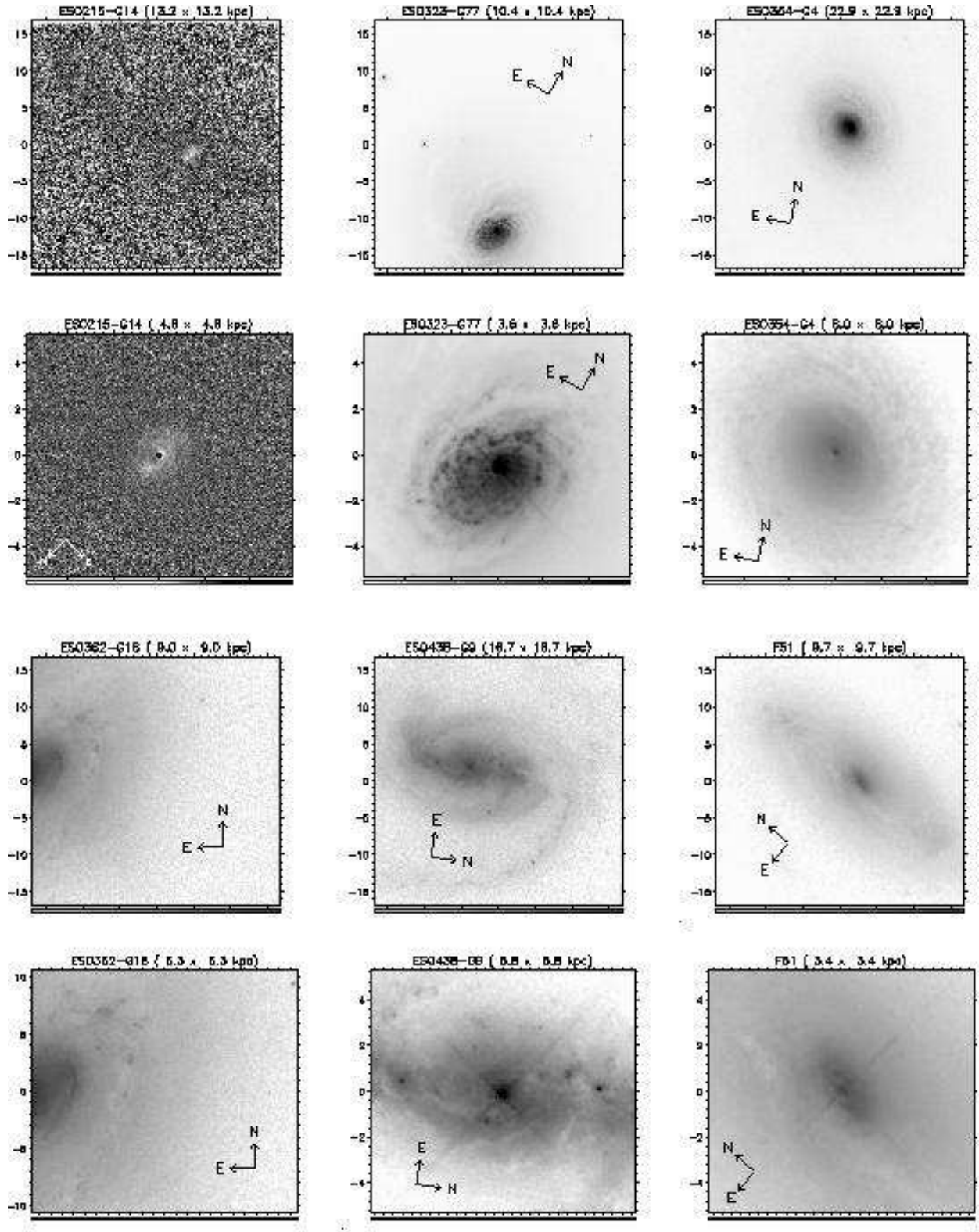}
  \caption{Structure maps of Seyferts 1s in Table~1}
\end{figure}
\clearpage
\begin{figure}[p]
  \setcounter{figure}{1}
  \centering
  \plotone{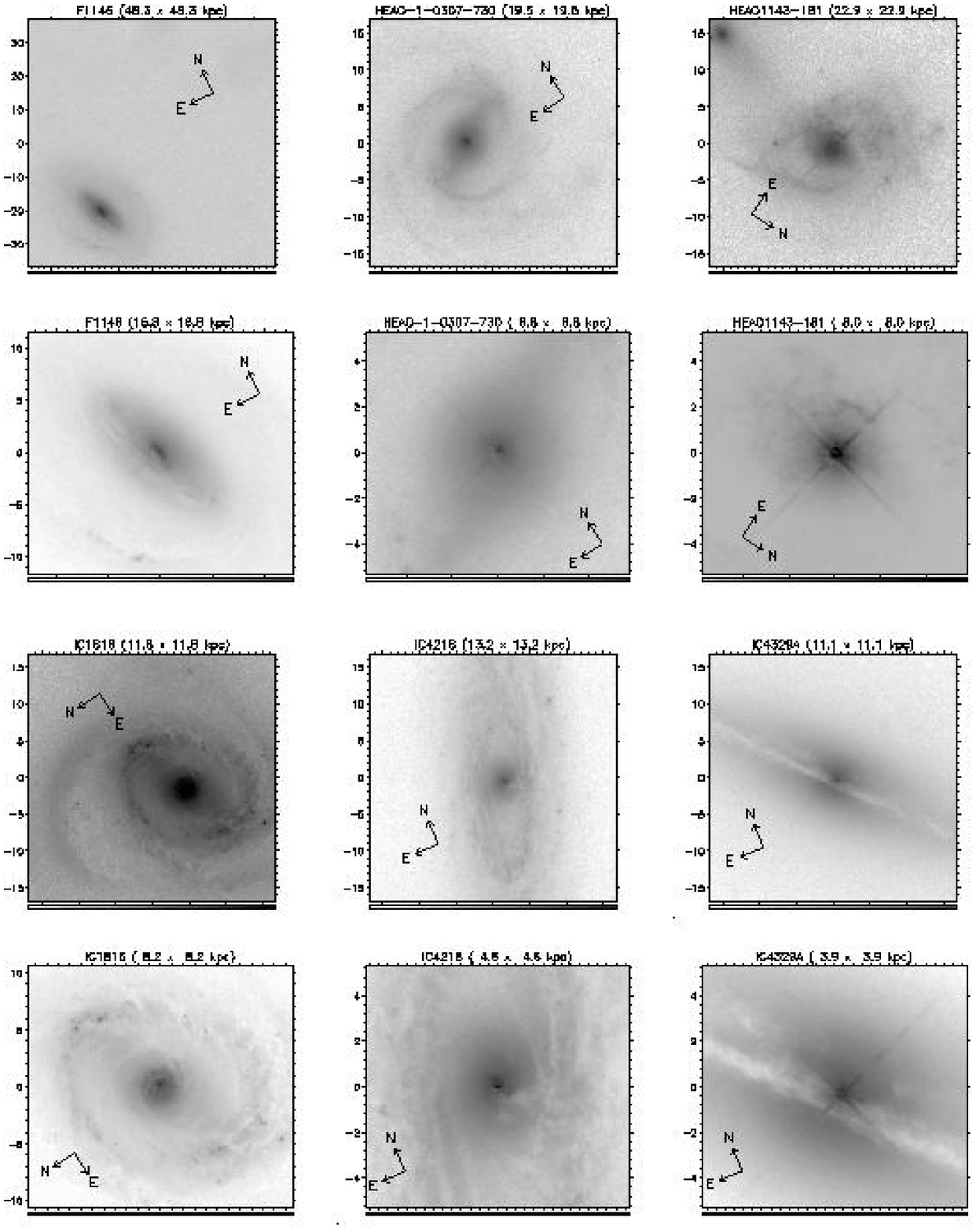}
  \caption{Continued}
\end{figure}
\clearpage
\begin{figure}[p]
  \setcounter{figure}{1}
  \centering
  \plotone{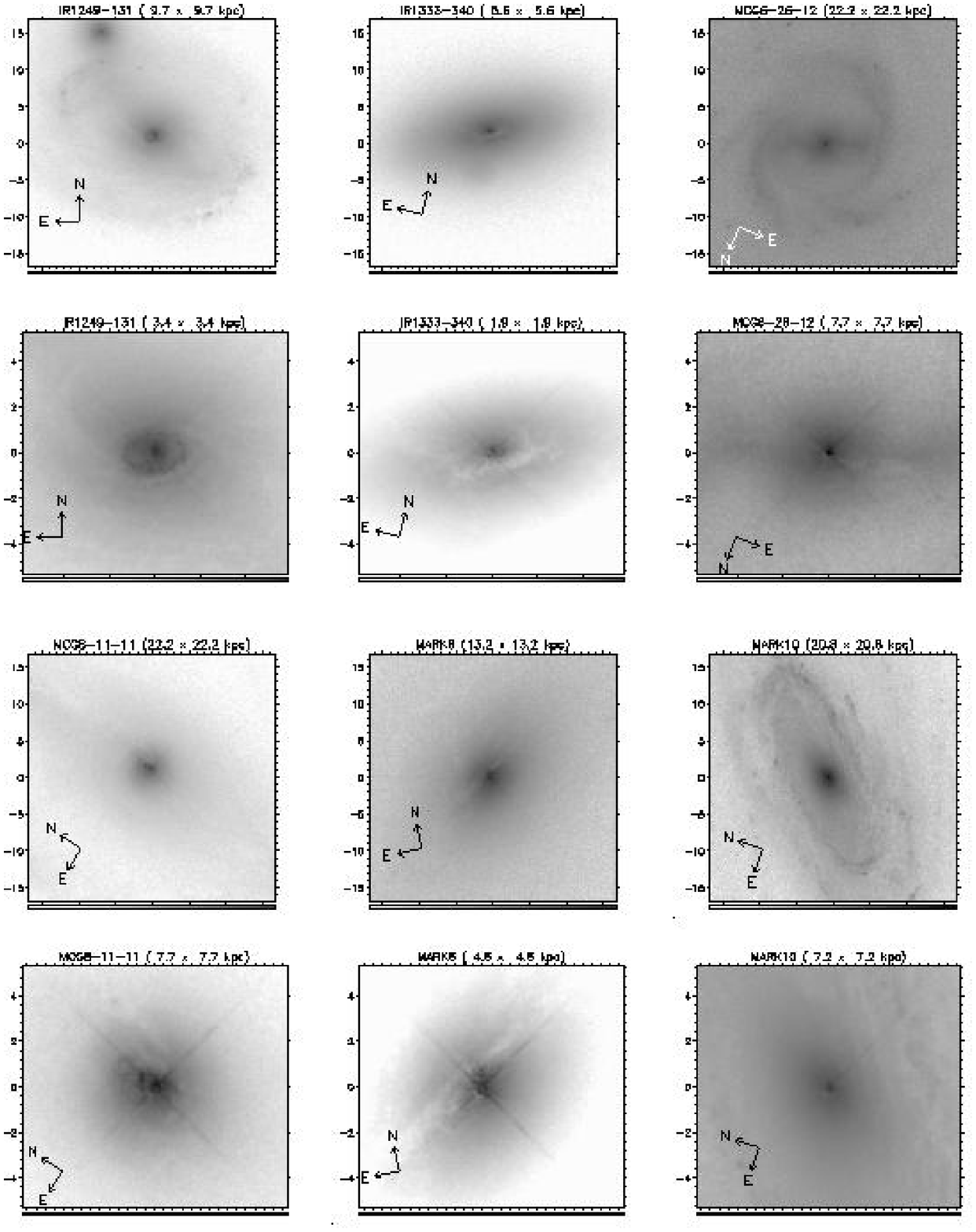}
  \caption{Continued}
\end{figure}
\clearpage
\begin{figure}[p]
  \setcounter{figure}{1}
  \centering
  \plotone{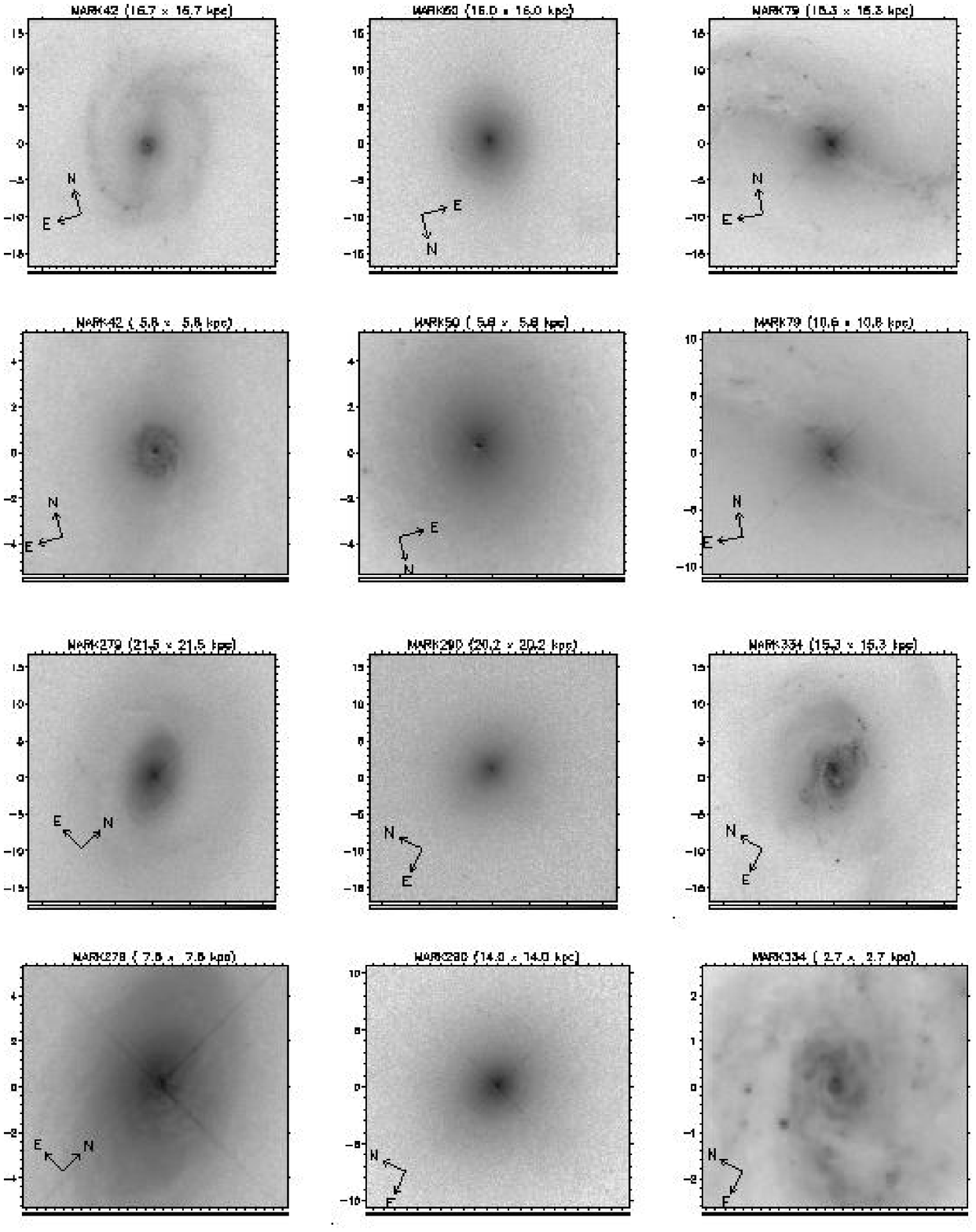}
  \caption{Continued}
\end{figure}
\clearpage
\begin{figure}[p]
  \setcounter{figure}{1}
  \centering
  \plotone{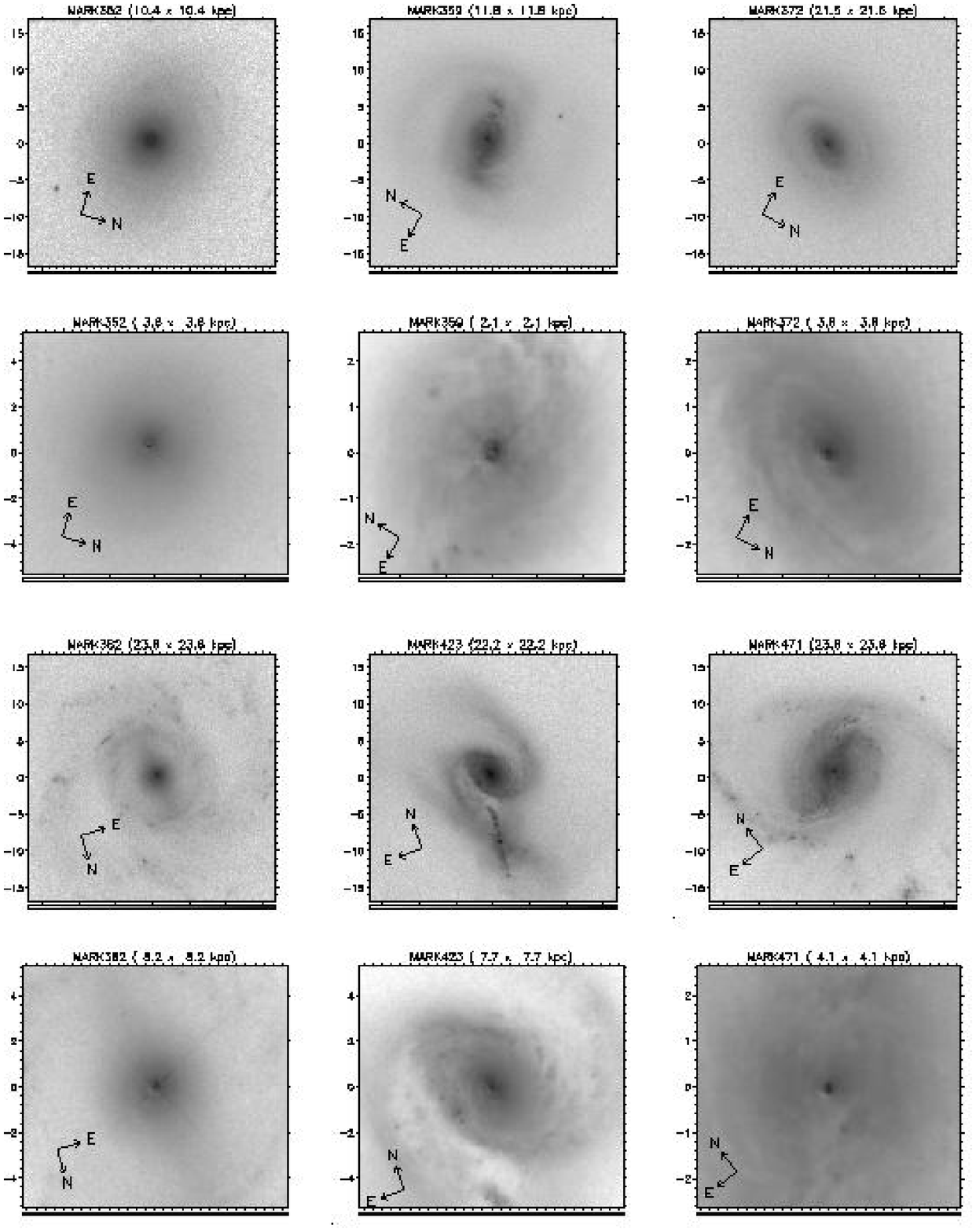}
  \caption{Continued}
\end{figure}
\clearpage
\begin{figure}[p]
  \setcounter{figure}{1}
  \centering
  \plotone{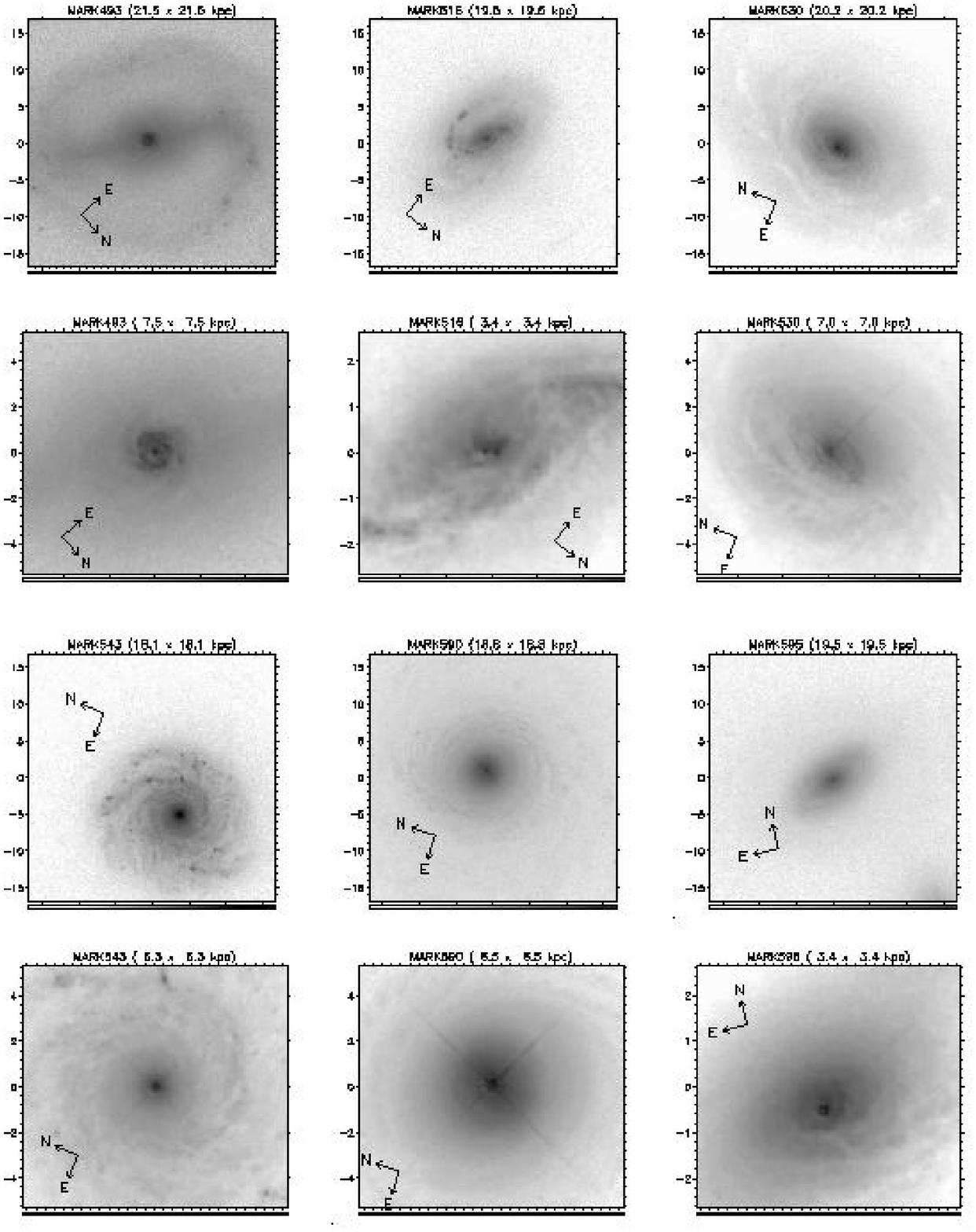}
  \caption{Continued}
\end{figure}
\clearpage
\begin{figure}[p]
  \setcounter{figure}{1}
  \centering
  \plotone{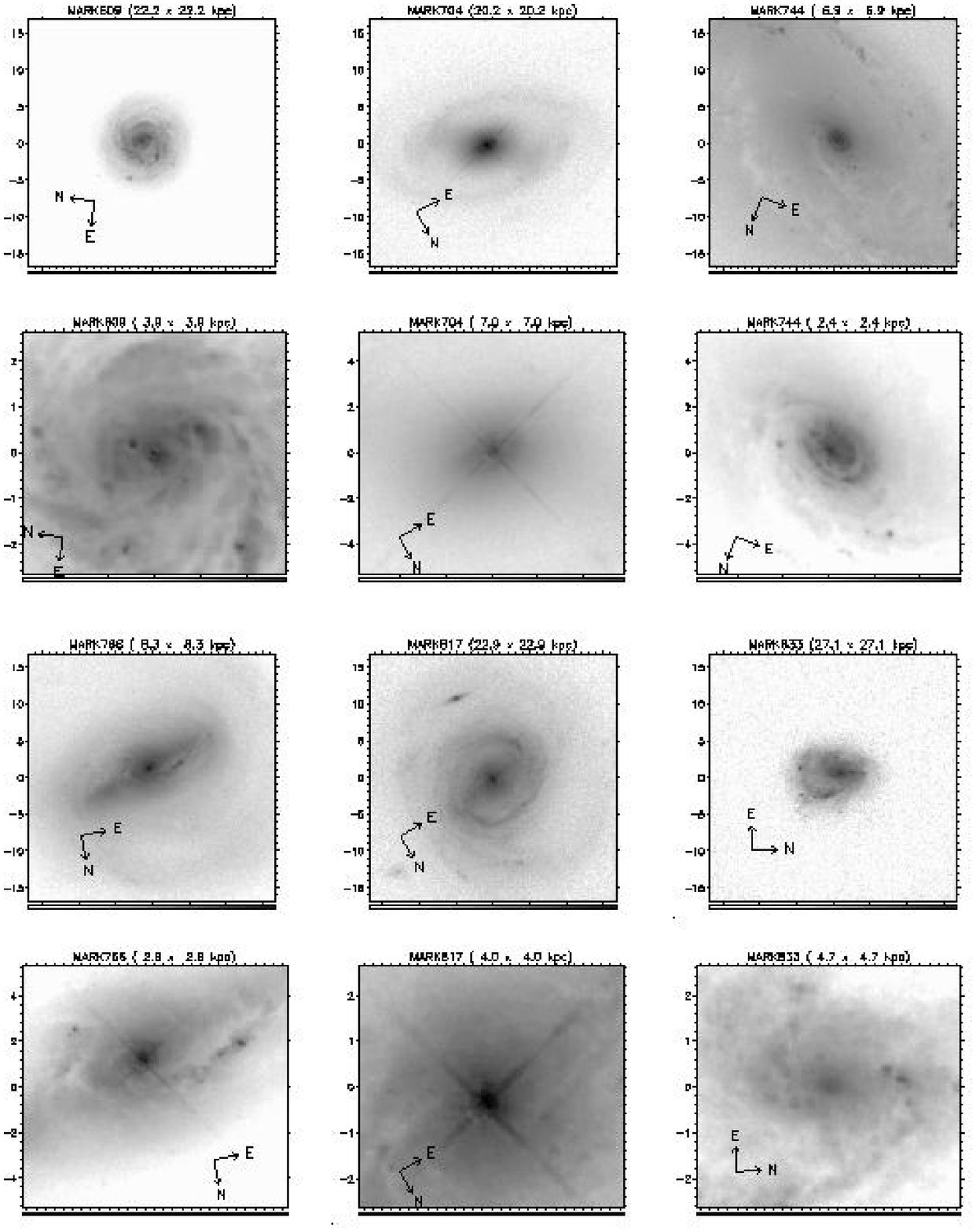}
  \caption{Continued}
\end{figure}
\clearpage
\begin{figure}[p]
  \setcounter{figure}{1}
  \centering
  \plotone{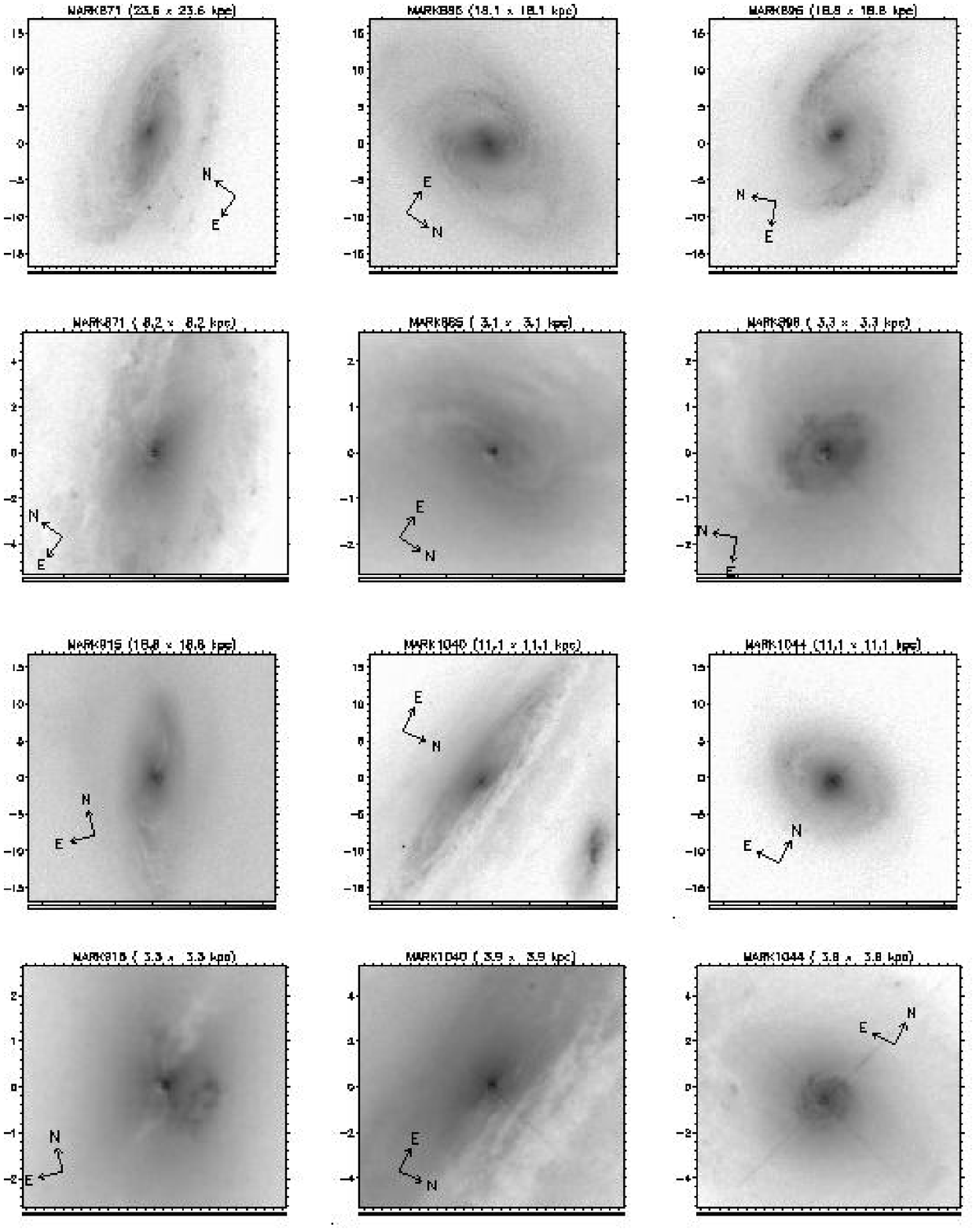}
  \caption{Continued}
\end{figure}
\clearpage
\begin{figure}[p]
  \setcounter{figure}{1}
  \centering
  \plotone{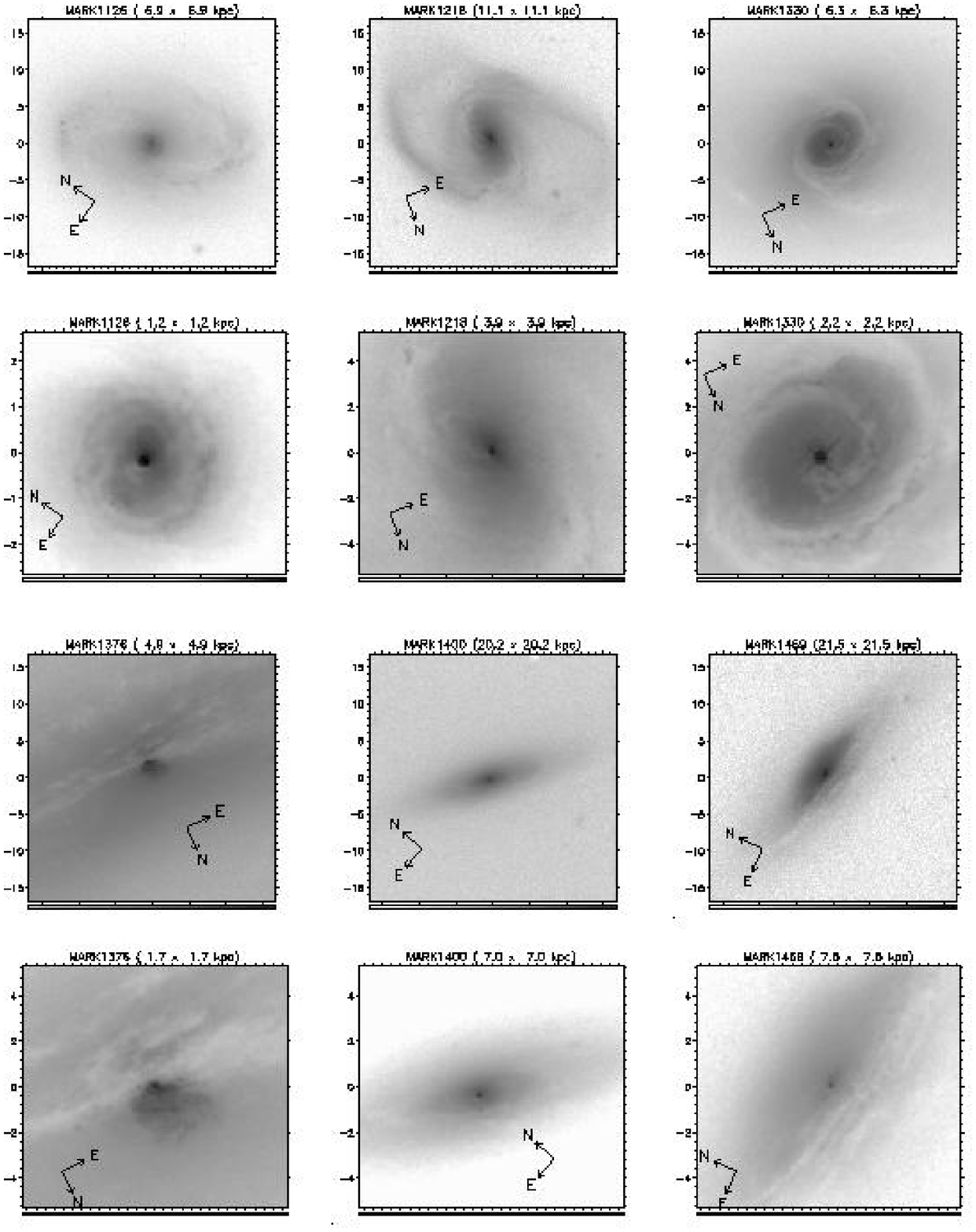}
  \caption{Continued}
\end{figure}
\clearpage
\begin{figure}[p]
  \setcounter{figure}{1}
  \centering
  \plotone{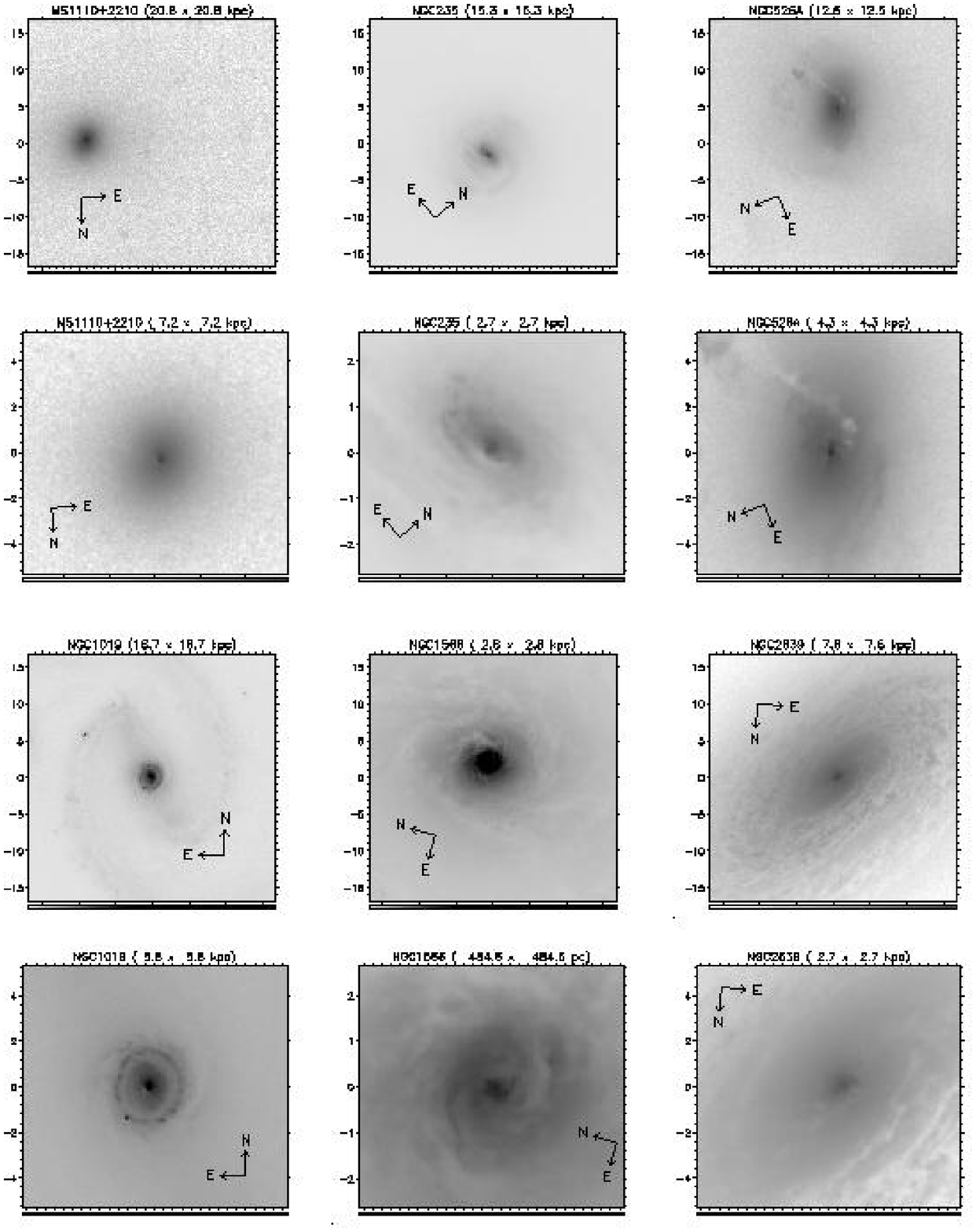}
  \caption{Continued}
\end{figure}
\clearpage
\begin{figure}[p]
  \setcounter{figure}{1}
  \centering
  \plotone{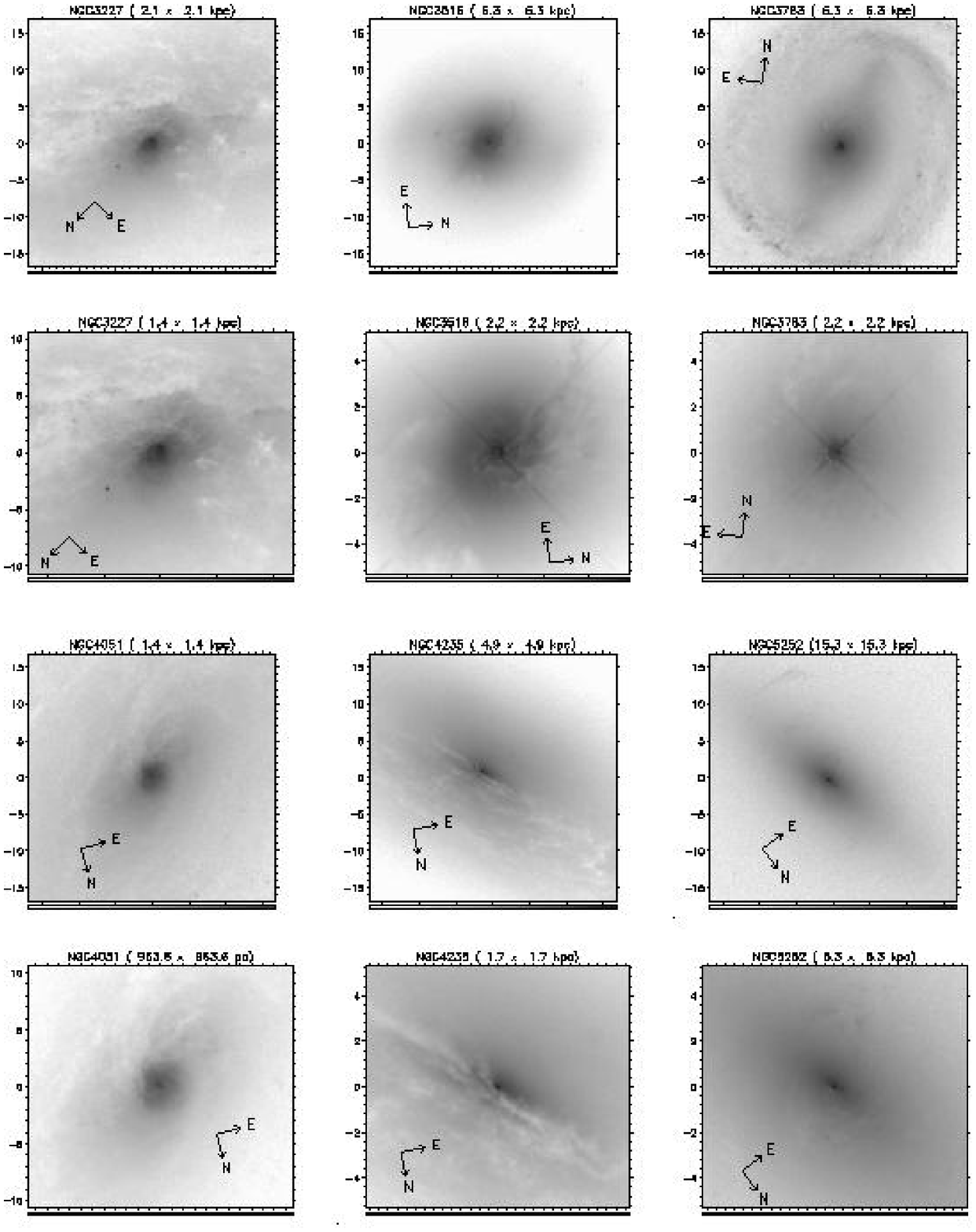}
  \caption{Continued}
\end{figure}
\clearpage
\begin{figure}[p]
  \setcounter{figure}{1}
  \centering
  \plotone{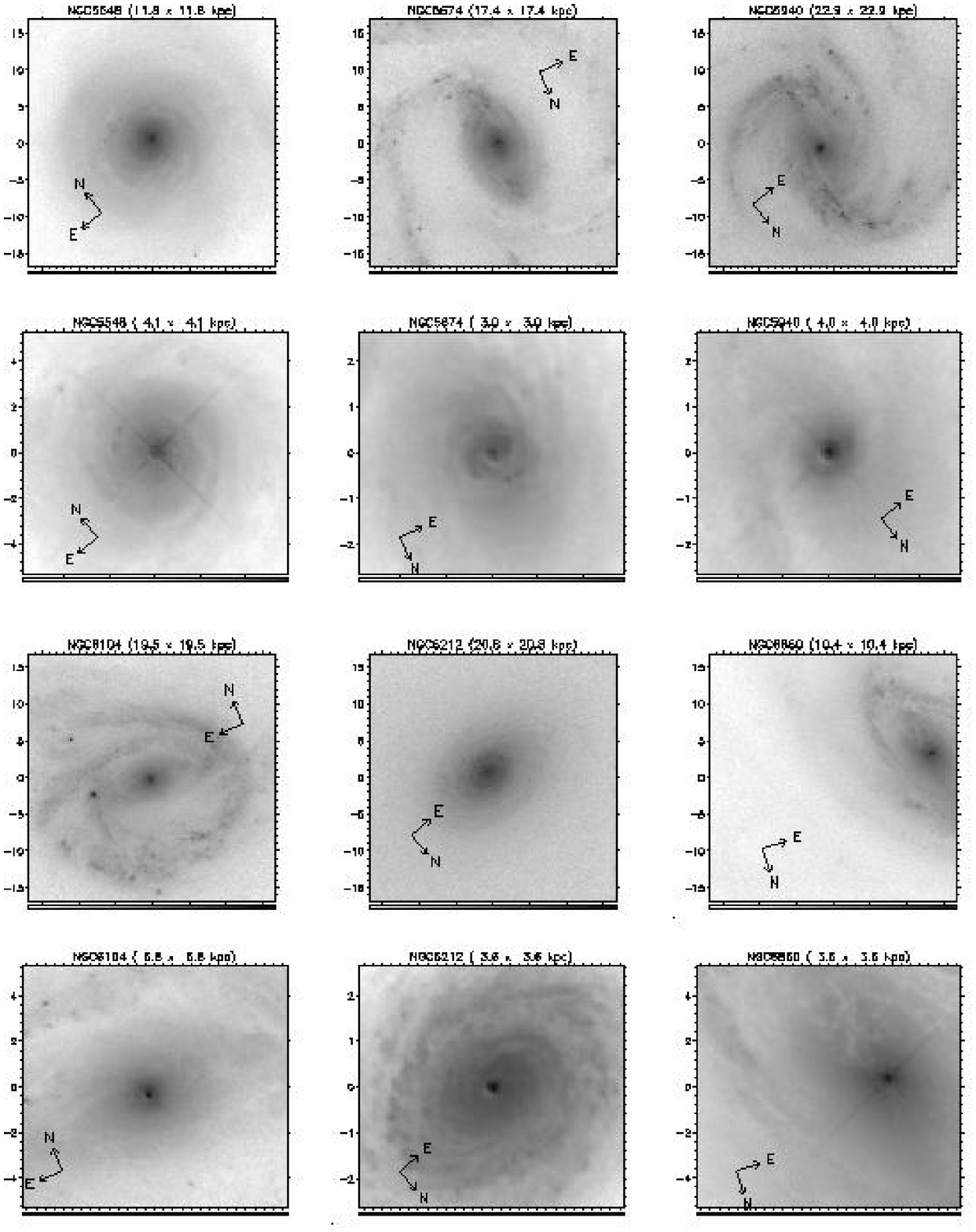}
  \caption{Continued}
\end{figure}
\clearpage
\begin{figure}[p]
  \setcounter{figure}{1}
  \centering
  \plotone{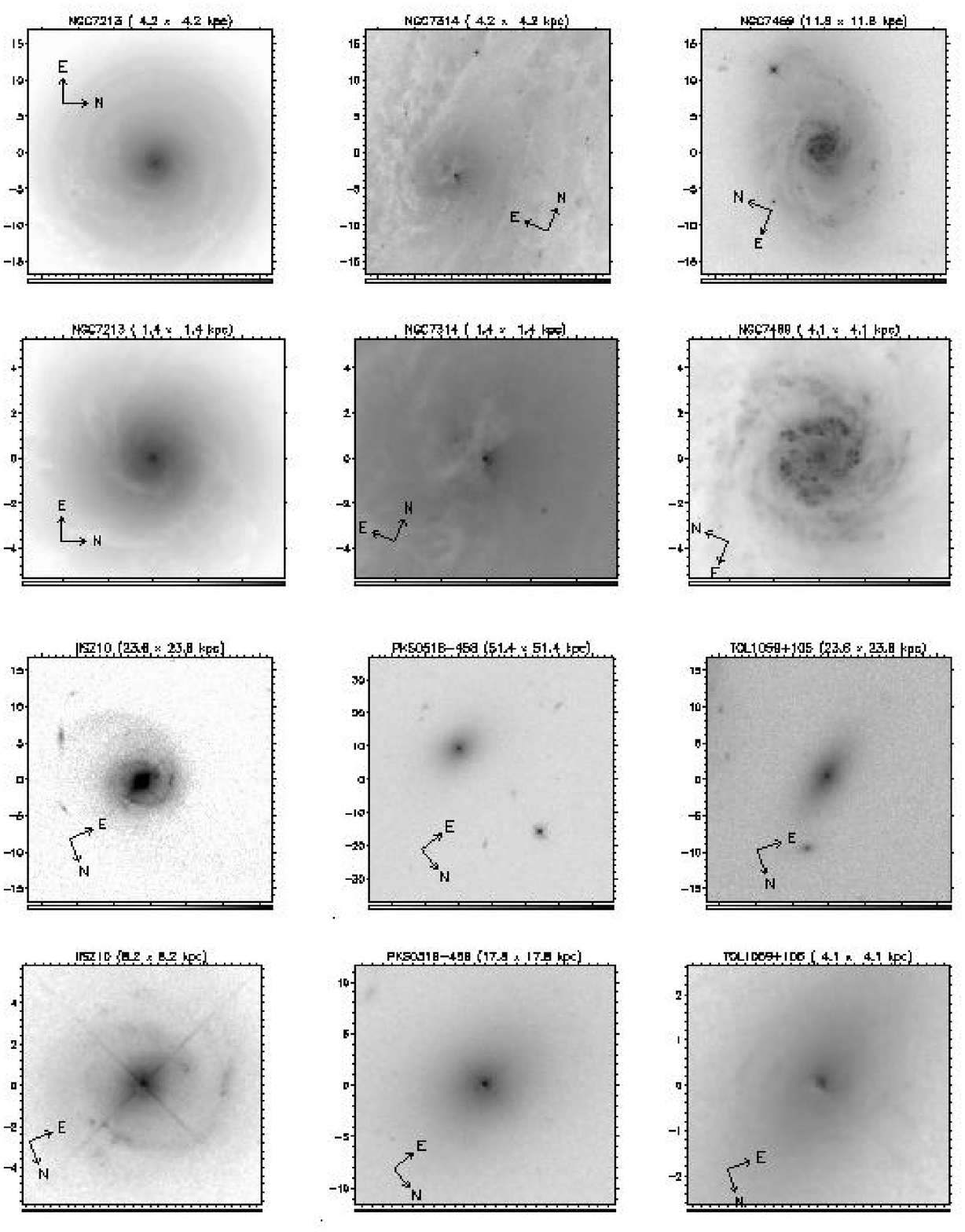}
  \caption{Continued}
\end{figure}
\clearpage
\begin{figure}[p]
  \setcounter{figure}{1}
  \centering
  \plotone{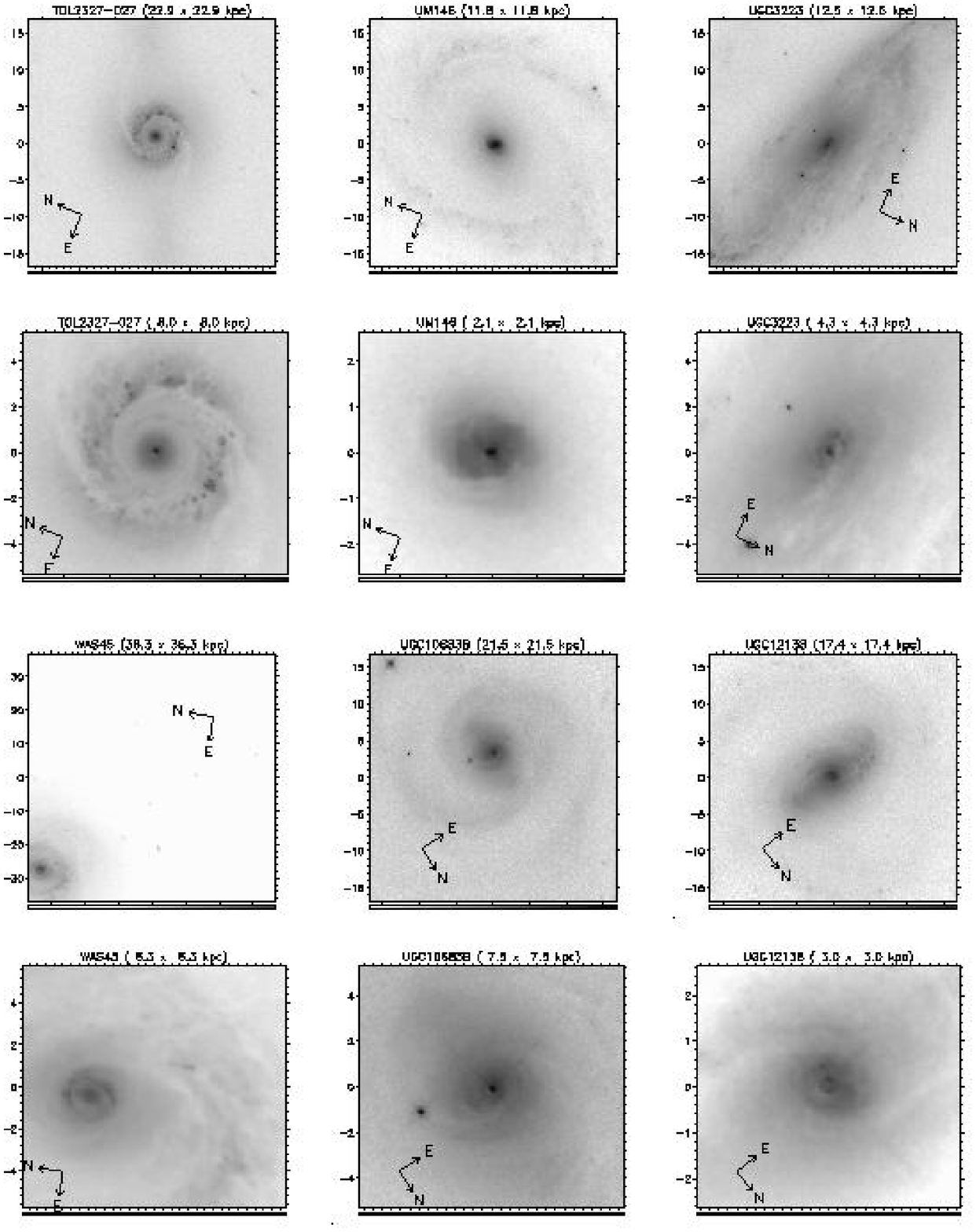}
  \caption{Continued}
\end{figure}
\clearpage
\begin{figure}[pt]
  \setcounter{figure}{1}
  \centering
  \plotone{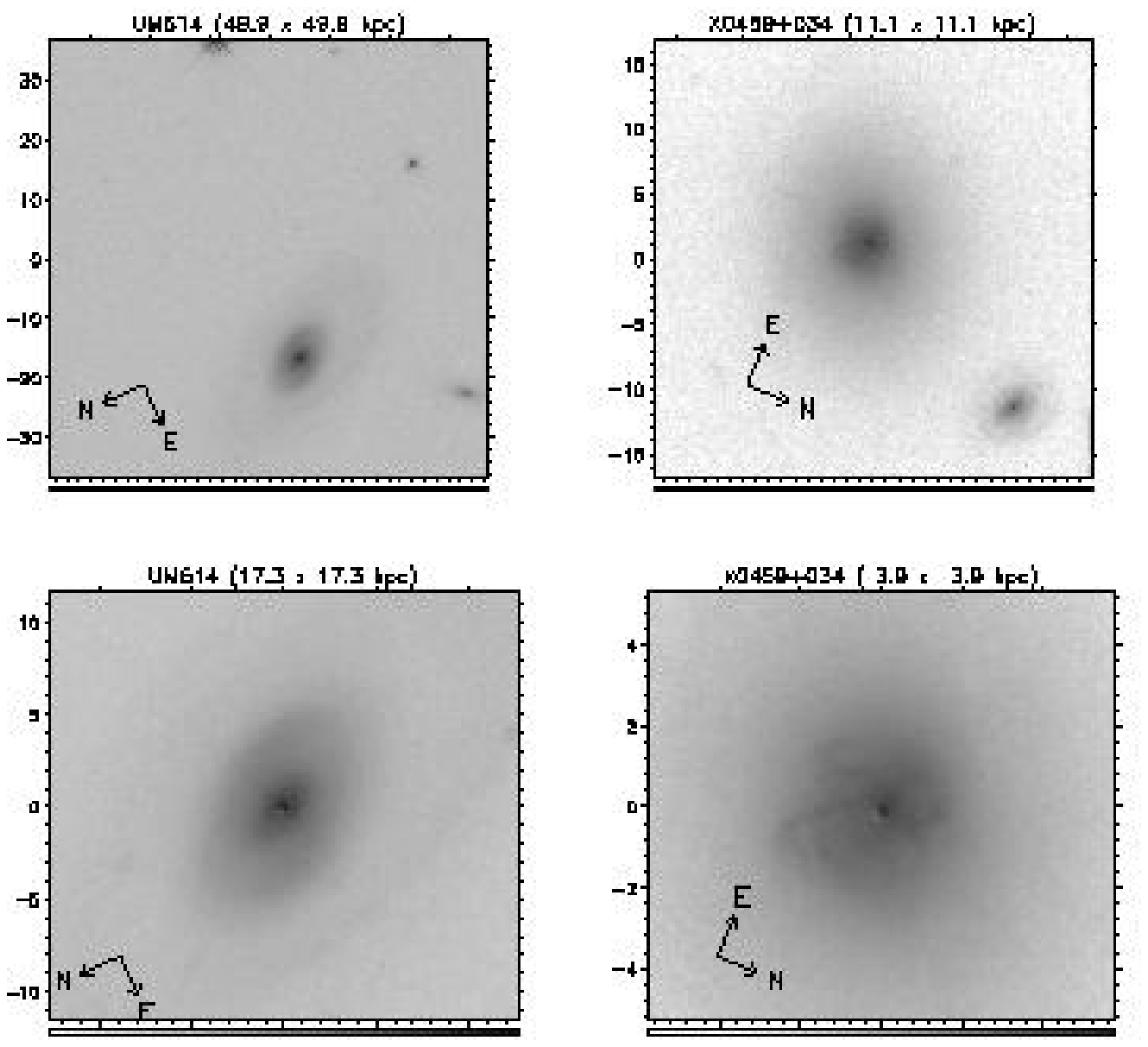}
  \caption{Continued}
\end{figure}
\clearpage

\newpage
%
%
\begin{figure}[p]
  \setcounter{figure}{2}
  \centering
  \plotone{fig3.eps}
  \caption{Frequency of Nuclear Dust Structures.}
\end{figure}

\end{document}